%% file: rms_atca_radio_new_v4.tex
\documentclass{aa}
\usepackage{amsmath}
\usepackage{amssymb}
\usepackage{graphicx}
\usepackage{epsfig}
\usepackage{changebar}
\usepackage{natbib}
\usepackage{multirow}
\usepackage{subfigure}
\usepackage{txfonts}

\usepackage{aalongtable}

\begin{document}
   \title{The RMS Survey:}

   \subtitle{Radio observations of candidate massive YSOs in the southern hemisphere\footnote{Tables 3 and 6 and full versions of Figs. 3, 7 and 8 are only available in electronic form at the CDS via anonymous ftp to cdsarc.u-strasbg.fr (130.79.125.5) or via http://cdsweb.u-strasbg.fr/cgi-bin/qcat?J/A+A/. Tables 5, 7 and 8 are only available as on-line material in the electronic edition of the journal at http://www.edpsciences.org.}}

   \author{J.~S.~Urquhart
          \inst{1}
          \and
			 A.~L.~Busfield
			 \inst{1}
			 \and
			 M.~G.~Hoare
			 \inst{1}
			 \and
			 S.~L.~Lumsden
			 \inst{1}
			 \and
			 A.~J.~Clarke
			 \inst{1}
			 \and
			 T.~J.~T.~Moore
			 \inst{2}
			 \and
			 J.~C.~Mottram
			 \inst{1}
			 \and
			 R.~D.~Oudmaijer
			 \inst{1}
          }

   \offprints{J. S. Urquhart: jsu@ast.leeds.ac.uk}

   \institute{School of Physics and Astrophysics, University of Leeds, Leeds, LS2~9JT, UK \\
         \and
             Astrophysics Research Institute, Liverpool John Moores University, 				Twelve Quays House, Egerton Wharf, Birkenhead, CH41~1LD, UK \\
             }

   \date{}

\abstract
   {The Red MSX Source (RMS) survey is a multi-wavelength programme of follow-up observations designed to distinguish between genuine massive young stellar objects (MYSOs) and other embedded or dusty objects, such as ultra compact (UC) HII regions, evolved stars and planetary nebulae (PNe). We have identified nearly 2000 MYSOs candidates by comparing the colours of MSX and 2MASS point sources to those of known MYSOs. }      
   {There are several other types of embedded or dust enshrouded objects that have similar colours as MYSOs and contaminate our sample. Two sources of contamination are from UCHII regions and PNe, both of which can be identified from the radio emission emitted by their ionised nebulae.} 
   {In order to identify UCHII regions and PNe that contaminate our sample we have conducted high resolution radio continuum observations at 3.6 and 6~cm of all southern MYSOs candidates ($235\degr< l < 350\degr$) using the Australia Telescope Compact Array (ATCA). These observations have a spatial resolution of $\sim$1--2\arcsec\ and typical image rms noise values of $\sim$0.3~mJy -- sensitive enough to detect a HII region powered by B0.5 star at the far side of the Galaxy.}
   {Of the 826 RMS sources observed we found 199 to be associated with radio emission, $\sim$25\% of the sample. The Galactic distribution, morphologies and spectral indices of the radio sources associated with the RMS sources are consistent with these sources being UCHII regions. Importantly, the 627 RMS sources for which no radio emission was detected are still potential MYSOs. In addition to the 802 RMS fields observed we present observations of a further 190 fields. These observations were made towards MSX sources that passed cuts in earlier versions of the survey, but were later excluded.}
  {}

\authorrunning{J. S. Urquhart et al.}
\titlerunning{Radio observations of MYSO candidates}
\maketitle
\section{Introduction}

During their relatively short lives massive O and early B type stars can have an enormous impact on their local environments. They inject huge amounts of energy into the interstellar medium (ISM) in the form of radiation and via molecular outflows, powerful stellar winds and supernova explosions. The radiation emitted is in the form of UV-photons which ionise their natal molecular cloud leading to the formation of an ultra compact (UC) HII region. Initially deeply embedded these expand, breaking free of their natal cloud and eventually evolving into the more classical HII region. The radiation also heats the surrounding material, evaporating ice mantles from the surfaces of dust particles, which alters the local chemistry. Throughout their lives massive stars process a huge amount of material and are responsible for the production of most of the heavy elements, which are returned to the ISM upon the stars' death, ejected in the subsequent supernova explosion. Through this process massive stars play an important role in the enrichment of the ISM.

Given that massive stars have such a profound impact, not only on their local environment, but also on a Galactic scale, understanding the environmental conditions and processes involved in their birth and the earliest stages of their evolution are of fundamental importance. However, although our understanding of the evolution of massive stars once they have emerged from their natal molecular cloud has vastly improved over the past couple of decades, their formation and early stages of evolution are still poorly understood. Many of the difficulties involved in the investigation of the formation of massive stars are implicit to the environmental conditions necessary for their formation. The initial stages of their formation and evolution take place deeply embedded within dense cores which are opaque to traditional optical and UV probes. Massive stars are known to form exclusively in clusters and are generally located much farther away than regions of low-mass star formation, making it difficult to attribute derived quantities to individual sources.  Massive stars are much rarer than their low-mass counterparts, this is compounded by the fact that massive young stellar objects (MYSOs) evolve  much more rapidly than in the case of low-mass stars, reducing the number available for study still further.

For these reasons, until relatively recently, the only catalogue of MYSOs
had been limited to 30 or so serendipitously detected  sources (\citealt{Henning1984}) which are mostly nearby and may not be representative of MYSOs in general. The situation has improved considerably in the last few years with studies of many new MYSO candidates using various selection criteria, all of which are based on IRAS colours and therefore biased towards bright,  isolated sources. For example  \citet{molinari1996} used ammonia observations of a sample of colour selected IRAS point sources;  \citet{walsh1997} searched for methanol masers toward 535 UCHII regions, identified from their IRAS colours following the \citet{wood1989a} selection criteria; \citet{sridharan2002} produced a catalogue of 69 candidate MYSOs, located between 1--10 kpc, selected on the basis of far-infrared, radio continuum and molecular line data. Many genuine MYSOs have been identified, however, due to the various ways these catalogues have been selected it is unclear whether these studies are typical of the general population of MYSOs, or whether the different selection criteria have resulted in the extraction of sub-groups of the general population. The lack of a large unbiased sample of MYSOs makes statistical studies of the more general population of MYSOs difficult.


MYSOs are mid-infrared bright with luminosities of 10$^4$--10$^5$ L$_\odot$ (\citealt{wynn-williams1982}). If, as expected, there is
a general trend from deeply to less embedded as the accretion and
dissipation of the molecular cloud proceeds then this would indicate that
the MYSO stage is somewhat later than the hot molecular core (HMC) stage, which is usually difficult to detect in the mid-IR (e.g., G29.96-0.02; \citealt{de_buizer2002}). However, there is likely to be significant overlap between the two classes. Many MYSOs have been associated with massive bipolar molecular outflows (e.g., \citealt{wu2004}) and  therefore accretion is still likely to be ongoing. Although nuclear fusion has almost certainly begun, the large accretion rates are thought to quench the ionising radiation, and in so doing impede the formation of a UCHII region (\citealt{forster2000}). MYSOs are also known to possess ionising stellar winds but are weak thermal radio sources ($\sim$1~mJy at 1 kpc; \citealt{hoare2002}). These objects can therefore be roughly parameterised as mid-IR bright and relatively radio continuum quiet. \citet{lumsden2002} compared colours of sources from the MSX and 2MASS point source catalogues (\citealt{egan2003} and \citealt{cutri2003} respectively) to those of known MYSOs to develop a colour selection criteria which has been used to produced an unbiased sample of approximately 2000 candidate MYSOs. Although the Spitzer survey of the Galactic plane (GLIMPSE; \citealt{benjamin2005}), has
dramatically improved spatial resolution and sensitivity over that of MSX, it is clear that the selection of most MYSOs could not be colour-selected from GLIMPSE
catalogues since the vast majority are highly saturated. The more restricted
$l$ and $b$ range of GLIMPSE also means that many regions
would be missed; in particular, the outer Galaxy.


Unfortunately, there are several other types of embedded, or dust enshrouded objects, that have similar colours to MYSOs which contaminate our sample, such as UCHII regions, evolved stars and planetary nebulae (PNe). However, these contaminating sources can be identified by incorporating information obtained from other wavelengths. We are currently involved in a multi-wavelength programme of follow-up observations known as the Red MSX Source (RMS) survey (\citealt{hoare2005}). These observations are designed to distinguish between genuine MYSOs and other embedded or dusty objects. These include
high resolution radio cm continuum observations to identify UCHII regions and PNe, molecular line observations to obtain kinematic distances (e.g., \citealt{busfield2006}) and bolometric luminosities, mid-IR imaging to identify genuine point sources and to obtain accurate positions, and near-IR spectroscopy (e.g., \citealt{clarke2006}) to distinguish between MYSOs and evolved stars.   Our ultimate aim is to produce a large unbiased sample of MYSOs ($\sim$~500) with complementary multi-wavelength data with which to study their properties. 

A major source of contamination is thought to come from UCHII regions and to a much lesser degree some reddened, dusty PNe, both of which can be identified from the free-free radio emission emitted from their ionised nebulae. Radio continuum observations are therefore an ideal way of identifying these kinds of radio loud sources from the relatively radio quiet MYSOs. In this paper we present the observational results of our southern hemisphere radio observations, which included all southern RMS sources (235\degr$<l<$350\degr) that have not been previously observed at high resolution ($\sim$2$^{\prime\prime}$). Since kinematic distances are needed before the physical properties of these objects can be determined, which are not currently available for all sources, we restrict our analysis to the observational data and a statistical discussion of the results. We postpone a detailed investigation of the physical properties of the radio sources identified until  our molecular line observations has been completed (Hoare et al. 2006, in prep.). 

In Sect.~\ref{sect:observations} we outline our source selection, observational and data reduction procedures. We present contour maps of all radio detections and tabulate their observational parameters in Sect.~\ref{sect:results}. We present our discussion in Sect.~\ref{sect:discussion}. In Sect.~\ref{sect:summary} we present a summary of the results and highlight our main findings.

\section{Observations and data reduction procedure}
\label{sect:observations}

\subsection{Source selection}

\begin{figure*}
\begin{center}
\includegraphics[width=0.95\textwidth]{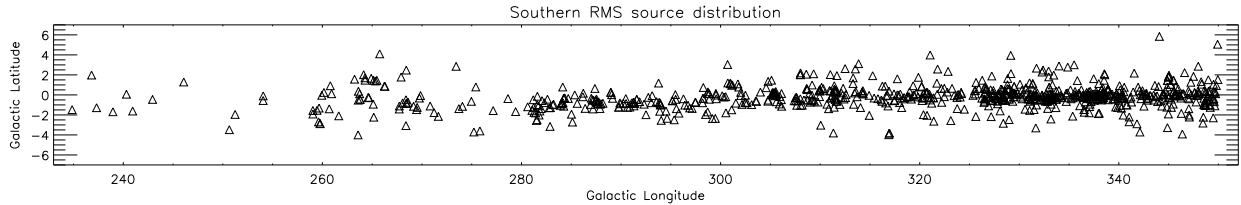}
\caption{The Galactic distribution of all southern hemisphere RMS sources located in the 3rd and 4th quadrants. This figure nicely illustrates the correlation between the distribution of RMS sources and the Galactic plane, and how their density increases towards the Galactic centre.}

\label{fig:rms_distribution}
\end{center}
\end{figure*}
 
Our colour selection criteria identified $\sim$2000 MYSO candidates spread throughout the Galaxy in a latitude range of $|b|<$5\degr. Sources toward the Galactic centre were excluded ($|l|<10\degr$) due to confusion and difficulties in calculating kinematic distances. Of these 892 are located  in the southern sky and are observable by the Australia Telescope Compact Array (ATCA)\footnote{The 
Australia Telescope Compact Array is funded by the Commonwealth of Australia for operation as a National 
Facility managed by CSIRO.}. The distribution of all southern RMS sources is presented in Fig.~\ref{fig:rms_distribution}.

In order to reduce the number of observations required, and to avoid observing sources that have been observed as part of other programmes, we took the following steps: 1) we cross-matched these source with previous high resolution surveys (e.g., \citealt{walsh1998}) eliminating sources for which high resolution data was already available; 2) conducted a literature search to identify and remove well known sources; 3) conducted a nearest-neighbour search to identify small groups of sources that were close enough to be observed in a single field. All of these procedures reduced the total number of sources that needed to be observed to 826, located within 802 fields.

\subsection{Observations}
\label{sect:rms_observations}

Radio continuum observations were made during four observing sessions between August 2002 and November 2004 using the ATCA, which is located at the Paul Wild Observatory, Narrabri, New South Wales, Australia. The ATCA consists of 6$\times$22 m antennas, 5 of which lie on a 3 km east-west
railway track with the sixth antenna located 3 km farther west. This allows the antennas to be positioned 
in several configurations with maximum and minimum baselines of 6 km and 30 metres respectively. Each antenna is fitted with a dual feedhorn 
system allowing simultaneous measurements at any two intermediate frequencies within the 3.6 and 6 cm bands.

\begin{table}
\begin{center}
\caption{Summary of observation dates, array configurations and numbers of sources observed.}
\label{tbl:observation_dates}
\begin{minipage}{\linewidth}
\begin{tabular}{ccccc}
\hline
\hline
Dates& Obs. 	& Arr.	& \multicolumn{2}{c}{Fields Observed} \\
	& time (h)  & config.&   RMS   & Additional\footnote{Observations towards sources that passed our initial colour selection criteria but were later excluded for various reasons (see Sect.~\ref{sect:additional_obs} for details).}	\\
\hline

25,28th/08/2002 & 14		& 6C	&  37  &22\\
19--22nd/07/2003 & 48		& 6D	&  152 &156 \\
15--19th/05/2004 & 60		& 6C	&  309 &1\\
24--30th/11/2004 & 72		& 6D	&  304 &7\\

\hline
\end{tabular}\\

\end{minipage}
\end{center}
\end{table}

The observations were made using a six km baseline configuration (either the 6C or 6D, see Table~\ref{tbl:observation_dates} for details) at 3.6 and 6~cm. We used a bandwidth of 128 MHz for each frequency with the bands centred at 4800 and 8640 MHz. Observations of each field typically consisted of a series of five 2 minute integration cuts (``snapshots''), spaced over a wide range of hour angles in order to maximise \emph{uv} coverage. The total on-source integration time of ten minutes gives a theoretical rms noise of $\sim$0.2~mJy at both frequencies, sufficient to detect a B0.5  or earlier type star embedded within an ionised nebula that is optically thin to radio emission at the far side of the Galaxy ($\ge$13~mJy at $\sim$20 kpc; \citealt{giveon2005a}). Observations were generally carried out over a 12 hour period which provides sparse but even coverage; however, for the first observing session the observations were carried out over a 5 hour period. The resulting synthesised beam for these observations are relatively poor and have elongated major axis. The observational parameters are summarised in Table~\ref{tbl:radio_setup}.

\begin{table}
\begin{center}
\caption{Summary of the observational parameters for the two frequencies used for our observations.}
\label{tbl:radio_setup}
\begin{minipage}{\linewidth}
\begin{tabular}{lcc}
\hline
\hline
Parameter& 3.6 cm & 6 cm \\
\hline
Rest frequency (MHz)\dotfill & 8640 & 4800 \\
Bandwidth (MHz)\dotfill& 128    & 128 \\
Primary beam \dotfill & 5$^{\prime}$.5 & 9$^{\prime}$.9 \\
Synthesised beam\footnote{Declination and hour angle dependent.}\dotfill   & $\sim$1$^{\prime\prime}$.5& $\sim$2$^{\prime\prime}$.5 \\
Largest well imaged structure$^a$\dotfill & $\sim$20$^{\prime\prime}$ &$\sim$30$^{\prime\prime}$\\
Theoretical image rms (mJy beam$^{-1}$)\dotfill & 0.21 & 0.22\\
Typical image rms (mJy beam$^{-1}$)\footnote{The stated rms values have been estimated from emission free regions close to the centre of maps presented in Fig.~3 (see Sect.~\ref{sect:results}).}\dotfill & 0.32 & 0.27 \\
Image pixel size\dotfill &  0$^{\prime\prime}$.33 & 0$^{\prime\prime}$.6\\

\hline
\end{tabular}\\

\end{minipage}
\end{center}
\end{table}

Sources were grouped by position into small blocks of between 8--10 sources, with each source being observed for 2 minutes within the block, and with each block being observed five times. To correct for fluctuations in the phase and 
amplitude of these data, caused by atmospheric and instrumental effects, each block was sandwiched between two short observations of a nearby phase calibrator (typically 2--3 minutes depending on the flux density of the calibrator). The primary flux 
calibrator 1934$-$638 was observed once during each set of observations to allow the absolute calibration of the flux density. The field names and positions are presented in Table~3 (available in electronic form at the CDS via anonymous ftp to cdsarc.u-strasbg.fr (130.79.125.5) or via http://cdsweb.u-strasbg.fr/cgi-bin/qcat?J/A+A/).

\subsection{Data reduction}
\label{sect:data_reduction}

The calibration and reduction of these data were performed using the MIRIAD reduction package \citep{sault1995} following standard ATCA procedures. We initially imaged a region equal to the size of the primary beam for each wavelength (i.e., 5.5$^{\prime}$ and 9.9$^{\prime}$ at 3.6 and 6 cm respectively). The image pixel size was chosen to provide $\sim$3 pixels across the synthesised beam for each wavelength (i.e., 0$^{\prime\prime}$.33 and 0$^{\prime\prime}$.6 for 3.6 and 6 cm respectively), resulting in image sizes of 1024$\times$1024 for both wavelengths. In a number of cases it was necessary to increase the size of the region being imaged due to the presence of compact bright sources on the edge or just outside the primary beam, which could be identified from their sidelobe pattern observed in the dirty map.

These maps were then deconvolved using a robust weighting of 0.5\footnote{A robust weighting of 0.5 produces images with the same sensitivity as natural weighting, but with a much improved beam-shape and lower sidelobe contamination.} and a couple of hundred cleaning components, or until the first negative component was encountered. These images were then examined for compact, high surface brightness sources using a nominal 4$\sigma$ detection threshold, where $\sigma$ refers to the image rms noise level. Fields in which radio emission was identified were subsequently re-imaged using between $\sim$300--1000 cleaning components in order to improve the dynamic range of the final maps. Finally the processed images were corrected for the attenuation of the primary beam.   

Due to the nature of interferometric observations the largest well-imaged structure possible at 3 and 6 cm, given our limited \emph{uv} coverage and integration time, is $\sim$ 20$^{\prime\prime}$ and 30$^{\prime\prime}$ respectively. However, many of our maps displayed evidence of large-scale emission, which, when under-sampled, can distort the processed images by producing image artifacts. These artifacts can appear as large undulations in the image intensity which are hard to remove and make identification of weak point sources embedded within extended regions of emission difficult (see \citealt{vla1999}, p. 127). Alternatively, the large-scale emission can become over-resolved and break up into irregular, or multiple component, sources that can often be confused with real sources. In order to limit the influence of large-scale emission, these fields were re-imaged, excluding the shortest baselines, and re-examined for high surface brightness sources (c.f. \citealt{kurtz1994}). The synthesised beam parameters of the reduced maps and an estimate of the maps rms noise level are presented in Table~3.

\section{Results}
\label{sect:results}
\subsection{RMS observational results}
\label{sect:obs_results}

We detected radio emission within a 12\arcsec\ radius (see later this section) of $\sim$25\% of the RMS sample observed (199 out of 826 RMS sources). The peak fluxes range from $\sim$1.2 mJy up to several hundred mJy, significantly above what would be expected from an ionising stellar wind ($\sim$1~mJy at 1~kpc; Hoare 2002). These sources are therefore unlikely to be genuine MYSOs but are most likely embedded UCHII regions. However, to avoid the possibility that a small number of these detections are actually due to ionised stellar winds in nearby sources we will, once the IR data becomes available, use the ratio of the radio and IR luminosities (i.e., Log L$_{\rm{radio}}$/L$_{\rm{IR}}$; see Fig.~6 of \citealt{hoare2005}) and retrospectively apply the following criteria: UCHII region $>$ 8, ionised stellar wind $<$ 8. Although this effectively eliminates a quarter of the RMS sources observed, the remaining 75\% (627 sources) towards which no radio emission was observed are still MYSO candidates.

In total we found 211 radio sources to be associated with the RMS sources, with multiple radio sources being associated with a single RMS sources in a few cases. All but seven of the radio sources detected towards the RMS sources were detected at 6~cm with the radio sources being detected at both frequencies towards 199 RMS sources. The positional offsets between the peak position of matched 3.6 and 6~cm radio sources were typically $\sim$ 0\arcsec.2. 


Ten RMS sources were found to be associated with two or more radio sources, possibly indicating a degree of clustering. The radio sources associated with these RMS sources are generally found in small compact groups with typical separations smaller than the resolution of the MSX beam ($<$18\arcsec), which has led to them being blended together into a single MSX point source. An example of a cluster of radio sources associated with an MSX sources can be seen in the \emph{lower right panel} of Fig.~\ref{fig:example_maps}. These multiple radio source matches could imply real clustering of UCHII regions. However, caution needs to be exercised when interpreting the multiplicity of these sources as it is also possible that the observed multiplicity could be due to  bright compact regions of emission located within a larger, more evolved HII region most of which is resolved out by the interferometer (e.g., \citealt{kurtz1994,thompson2006}). The resolution of the MSX data is too low to determine the true nature of these sources, but our high resolution mid-IR data will be able to distinguish between the two possibilities.
 
\begin{figure}
\begin{center}
\includegraphics[width=0.45\textwidth]{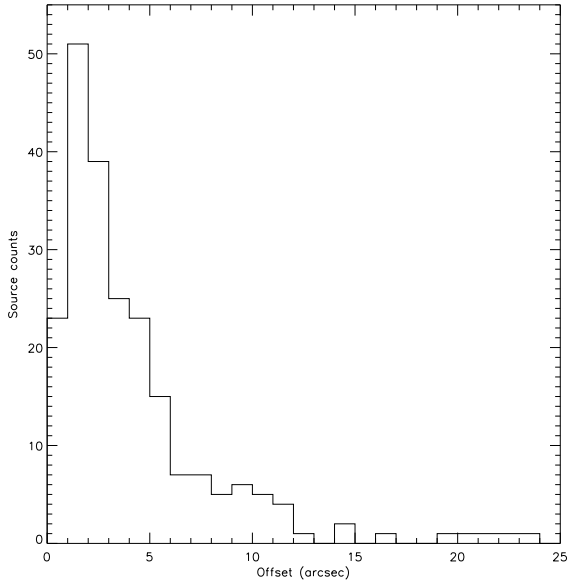}
\caption{Histogram of the projected angular separations between the detected radio sources and their nearest RMS point source counterpart.}

\label{fig:rms_radio_offsets}
\end{center}
\end{figure}

In Fig.~\ref{fig:rms_radio_offsets} we present a histogram of angular separation between the radio sources and their nearest RMS  source. The distribution of the angular separation between the radio and RMS source matches illustrates the tight correlation between the two. The distribution is strongly peaked at $\sim$1--2\arcsec\ and has a mean of $\sim$4--5\arcsec. The typical separations are $\sim$3\arcsec, which is comparable to the positional accuracy of the MSX point source catalogue ($\sim$2\arcsec; \citealt{egan1999}). The distribution peaks sharply and then falls steeply to $\sim$7\arcsec\ and then falls off more slowly until $\sim$12\arcsec\, after which the distribution flattens out to an almost constant level.

In this plot we use a rather generous 25\arcsec\ radius with which to look for possible associations between the radio detections, however, the distribution tails off after $\sim$12\arcsec\ to a constant background level which we consider marks the transition between real associations and chance alignments. Therefore we have chosen 12\arcsec\ as the cutoff radius for reliable associations of radio and RMS sources. Applying this radial cutoff we found over 90\% of all matched sources are located within 12\arcsec\ of each other, and more than 80\% are within 7\arcsec. In order to check if our cutoff radius was giving reliable associations we followed the procedures described by \citet{giveon2005a} for identifying radio-infrared matches. This analysis resulted in a reliability of 97\% or better for every RMS source matched with a radio source within a 12\arcsec\ radius, with the reliability falling off significantly for larger separations.

\subsection{Radio emission maps and source morphologies}
\label{sect:emission_maps}
\begin{figure*}
\begin{center}

\includegraphics[width=0.35\linewidth]{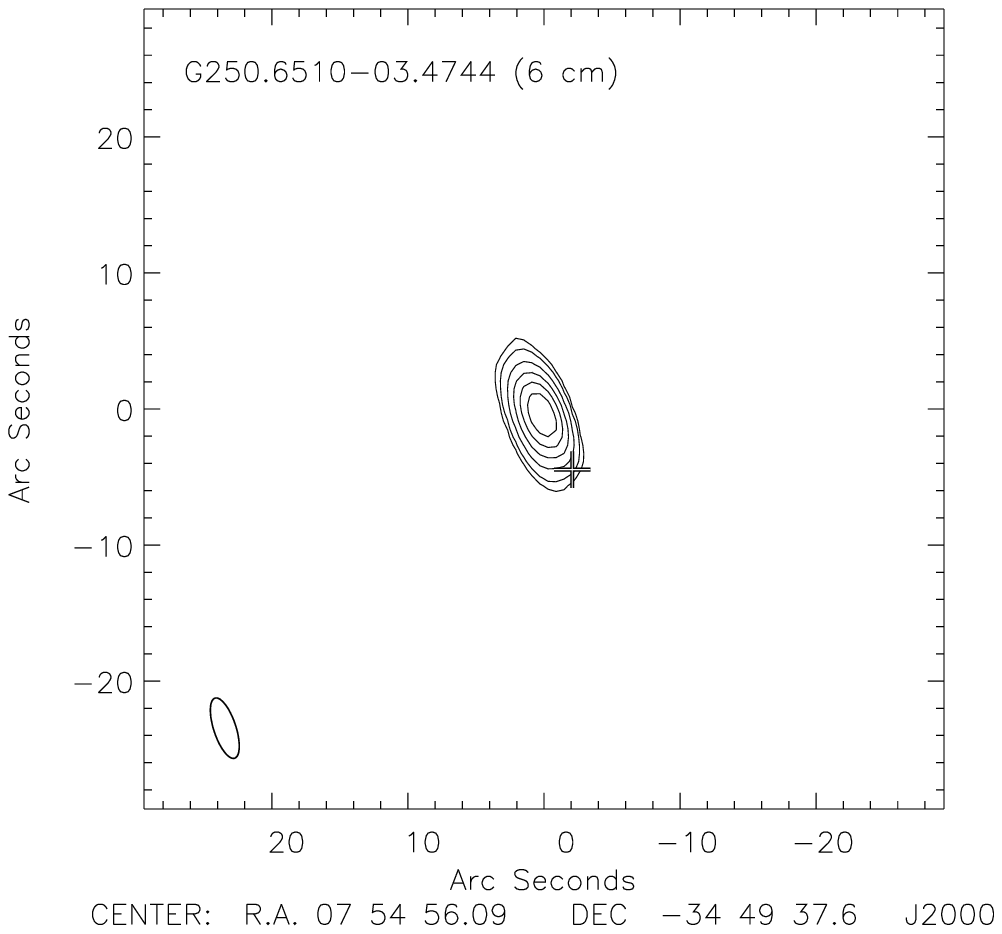} 
\includegraphics[width=0.35\linewidth]{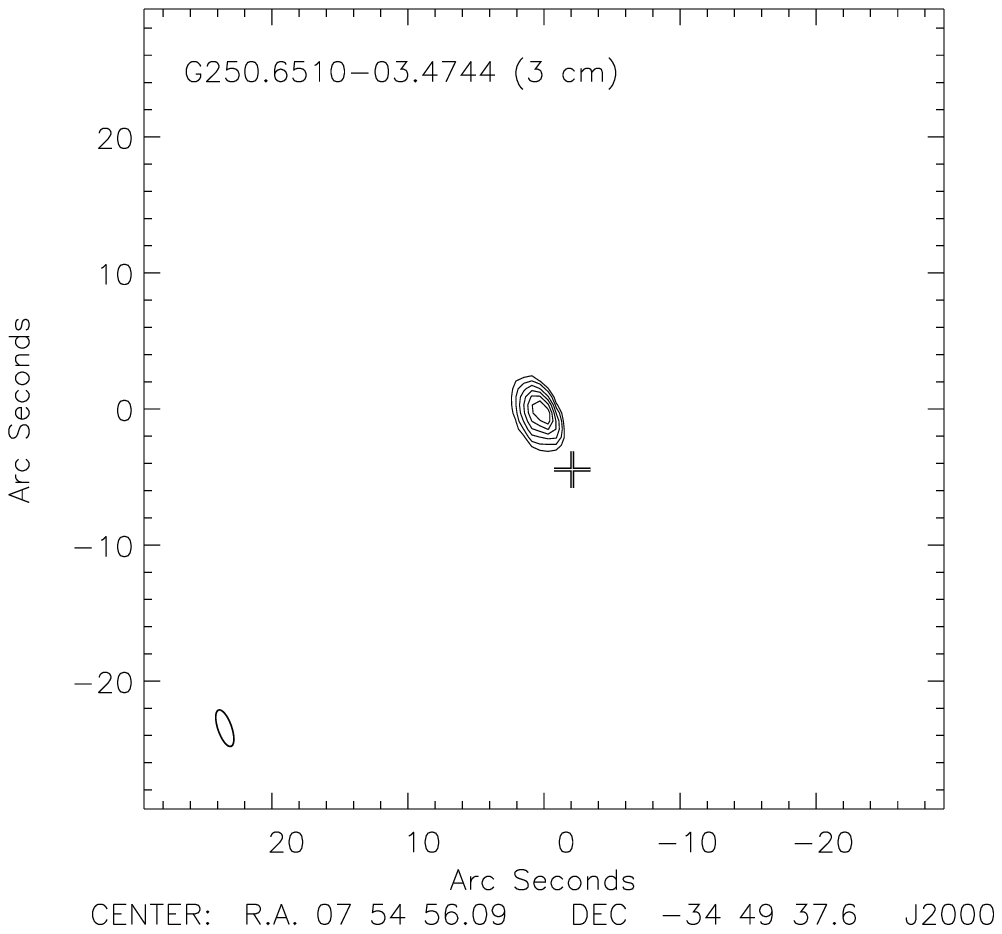}\\ 
\includegraphics[width=0.35\linewidth]{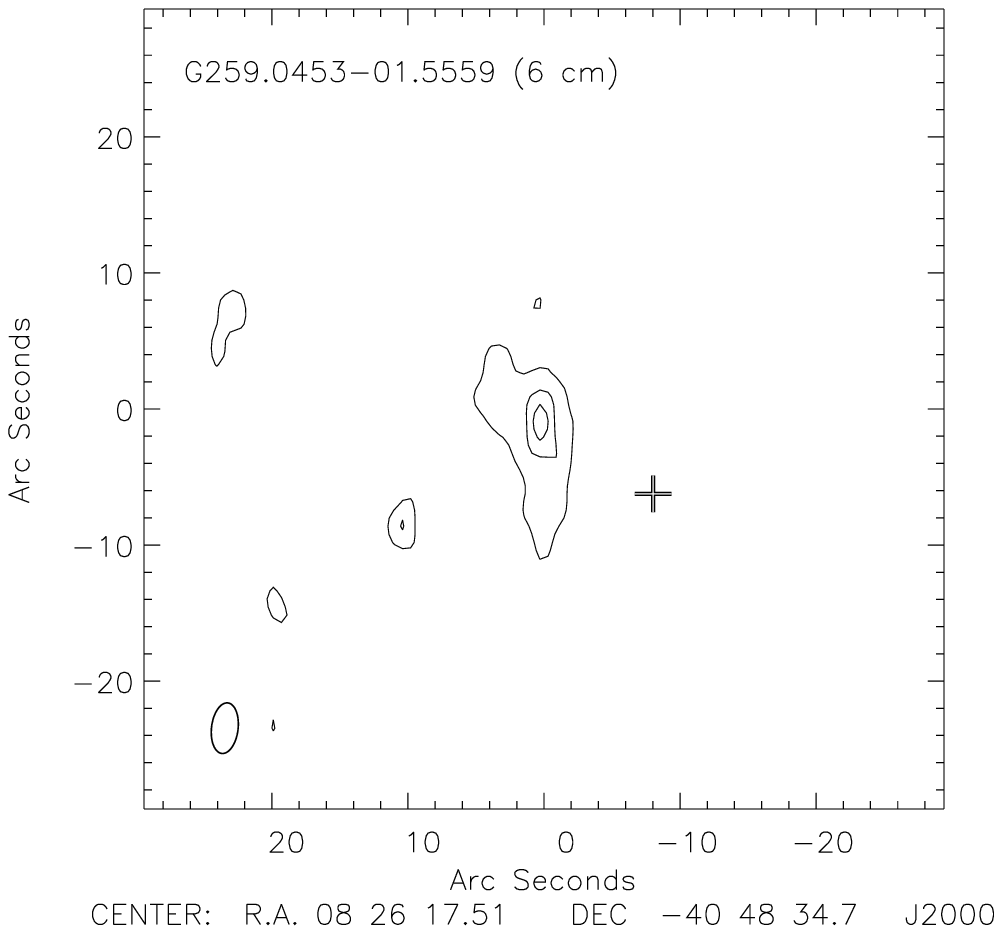}\\ 
\includegraphics[width=0.35\linewidth]{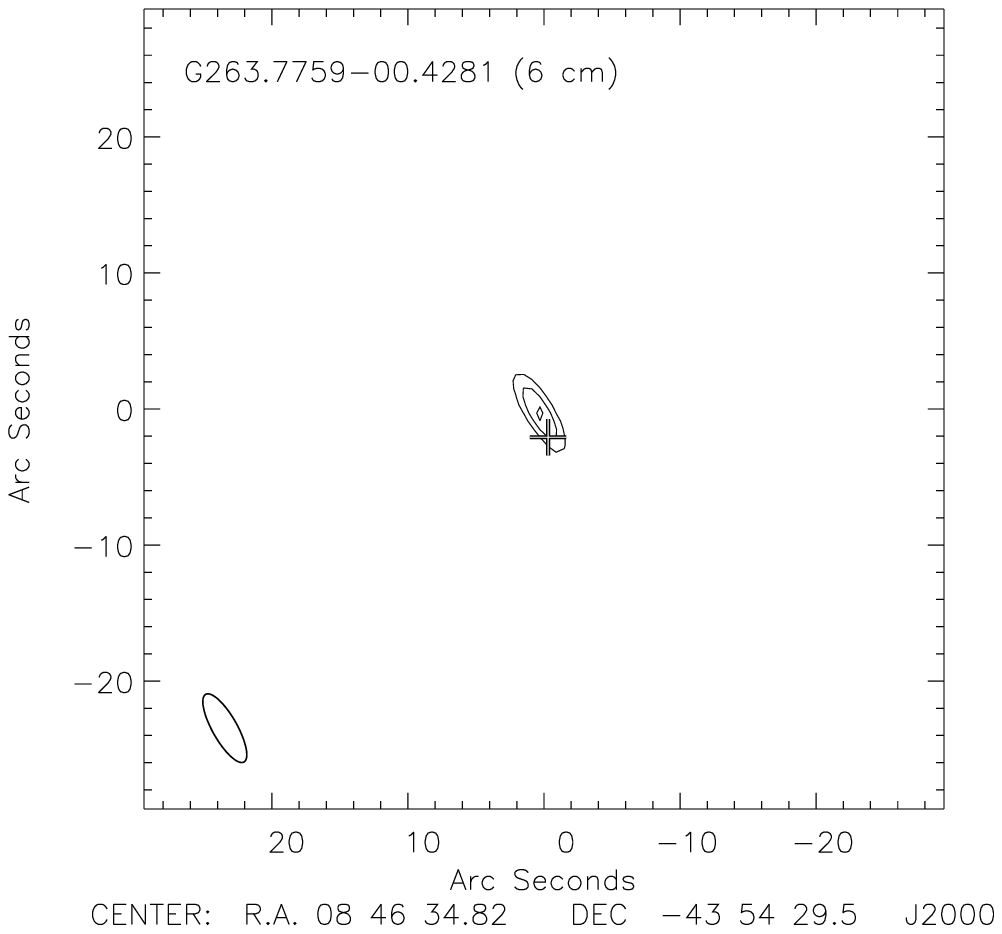} 
\caption{\label{fig:rms_maps} Contour radio maps of the radio sources with RMS counterparts. The first contour starts at 4$\sigma$ with the intervening levels determined by a dynamic power law (see text for details). The RMS name and wavelength are given in the top left corner. The size of the synthesised beam is shown to scale in the lower left hand corner. The full version of this figure is only available in electronic form at the CDS via anonymous ftp to cdsarc.u-strasbg.fr (130.79.125.5) or via http://cdsweb.u-strasbg.fr/cgi-bin/qcat?J/A+A/.}

\end{center}
\end{figure*}

In Fig.~\ref{fig:rms_maps} we present continuum maps of all RMS sources with associated radio emission in the form of contour plots. These plots cover a region of 1 arcmin$^2$ in size and are centred on the position of the RMS source. The RMS source name and the wavelength are given in the upper left corner of each plot. The position of the RMS source is indicated by a cross and the size of the synthesised beam is shown to scale in the lower left hand corner. These plots are presented in order of increasing galactic longitude. Radio sources that have been detected at both 3.6 and 6~cm are placed side by side (wavelength increasing to the right), cover the same field of view and are plotted using the same scale. Sources detected at a single frequency are placed in the centre of the page. In cases where multiple radio sources are associated with a single RMS source we have labelled each radio source alphabetically in order of increasing radial offset.

The contour levels in each map were determined using the dynamic range power-law fitting scheme described by \cite{thompson2006}. The advantage of this scheme over a linear scheme is in its ability to emphasise both emission from diffuse  extended structures with low surface brightness and emission from bright compact sources. The contour levels were determined using the following relationship $D=3\times N^i+4$, where $D$ is the dynamic range of the map (defined as the peak brightness divided by the map's rms noise), $N$ is the number of contours used (6 in this case), and $i$ is the contour power-law index. Note this relationship has been altered slightly from the one presented by  \citet{thompson2006} so that the first contours start at 4$\sigma$ rather than 3$\sigma$ used by them. The lowest power-law index used was one, which resulted in linearly spaced contour starting at 4$\sigma$ and increasing in steps of 3$\sigma$.


\begin{figure}
\begin{center}
~
\includegraphics[width=0.45\textwidth]{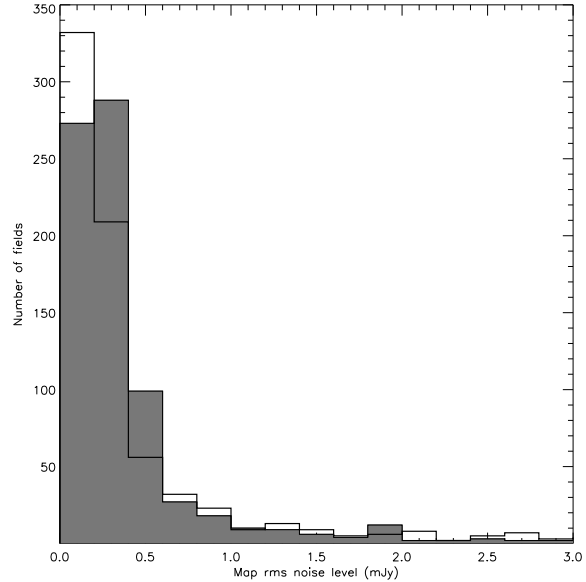}

\caption{\label{fig:completeness} Histogram of the fields observed as a function of map rms noise level. The 3.6~cm data is plotted as a filled histogram and the 6~cm data is over-plotted as a unfilled histogram.}

\end{center}
\end{figure}

In Fig.~\ref{fig:completeness} we present a plot of the number of fields as a function of the their map rms noise level for both the 3.6 and 6~cm observations. We plot the 6 cm data as an unfilled histogram and the 3.6~cm data as a filled histogram (grey). We truncate the x-axis at 3 mJy as the distribution tails off after this point, however, approximately 10\% of 6~cm maps and 6\% of 3.6~cm maps have rms noise levels above 3~mJy. The noise levels in both wavelength maps are similar with typical values of $\sim$0.3 mJy, only slightly worse than the theoretical noise level of $\sim$0.2 mJy. 

\begin{figure*}
\begin{center}

\includegraphics[width=0.35\linewidth]{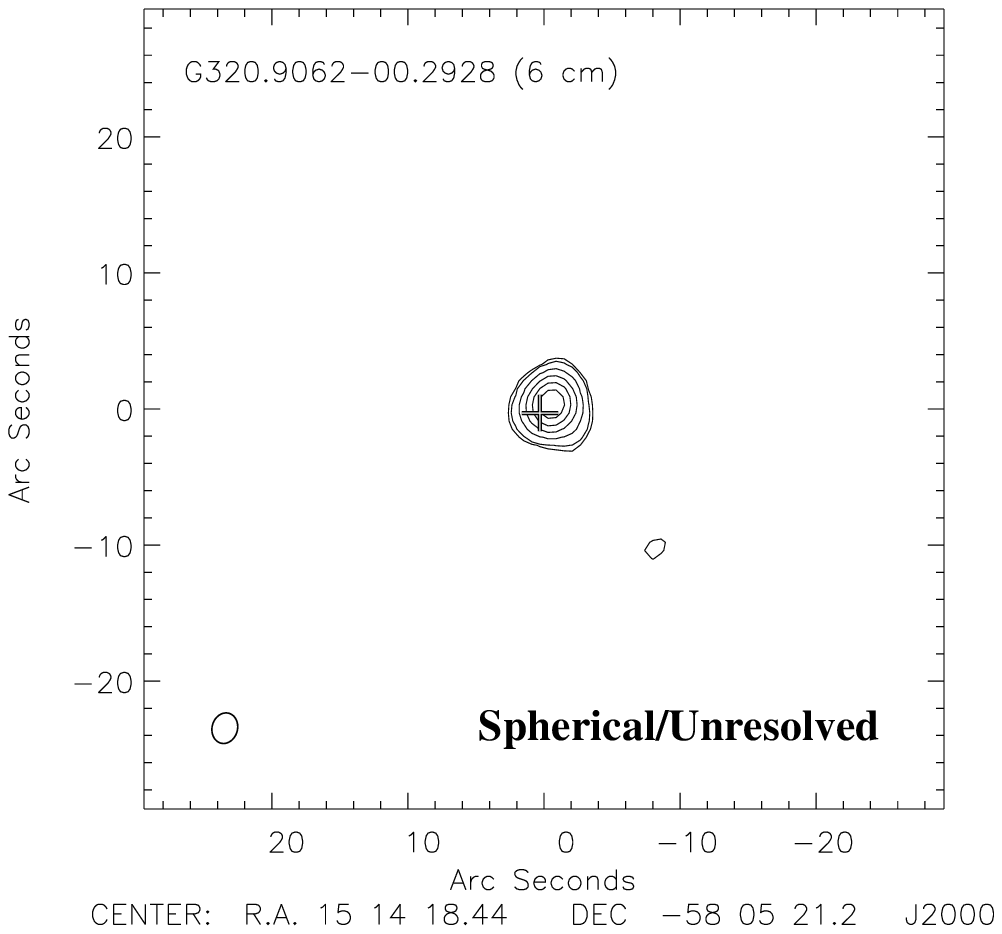} 
\includegraphics[width=0.35\linewidth]{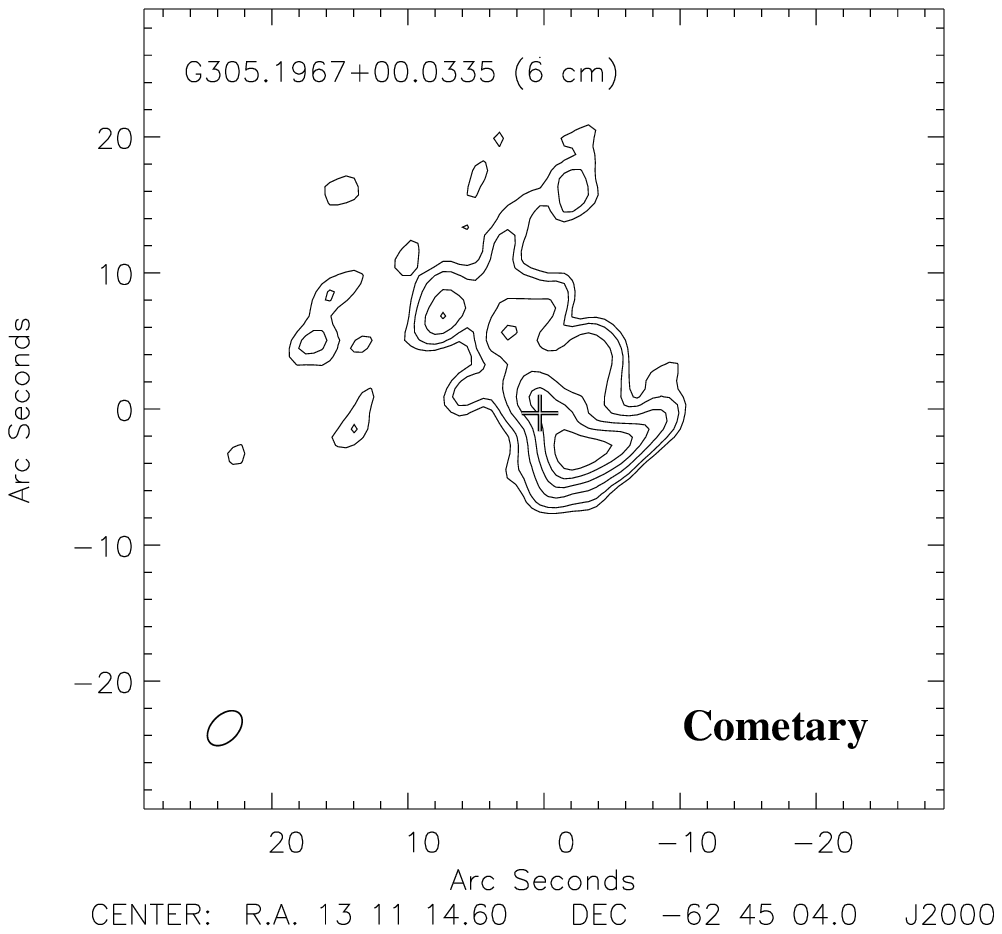}\\ 
\includegraphics[width=0.35\linewidth]{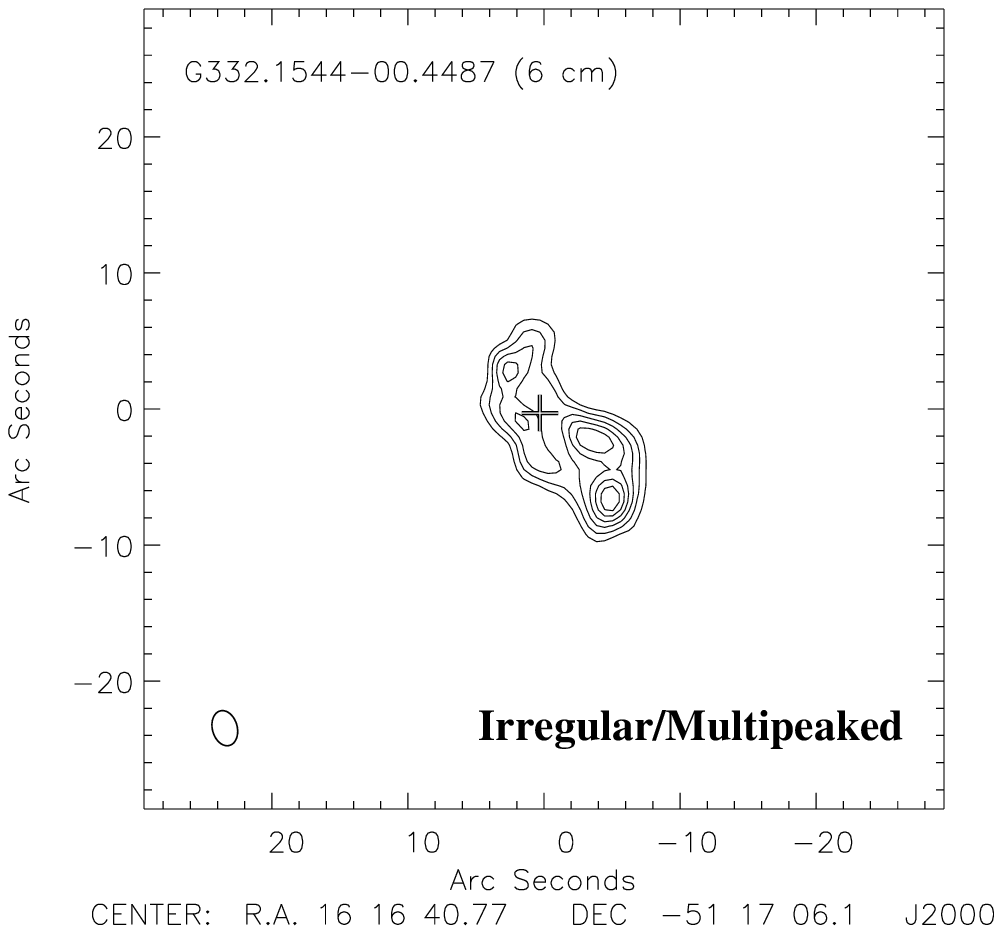} 
\includegraphics[width=0.35\linewidth]{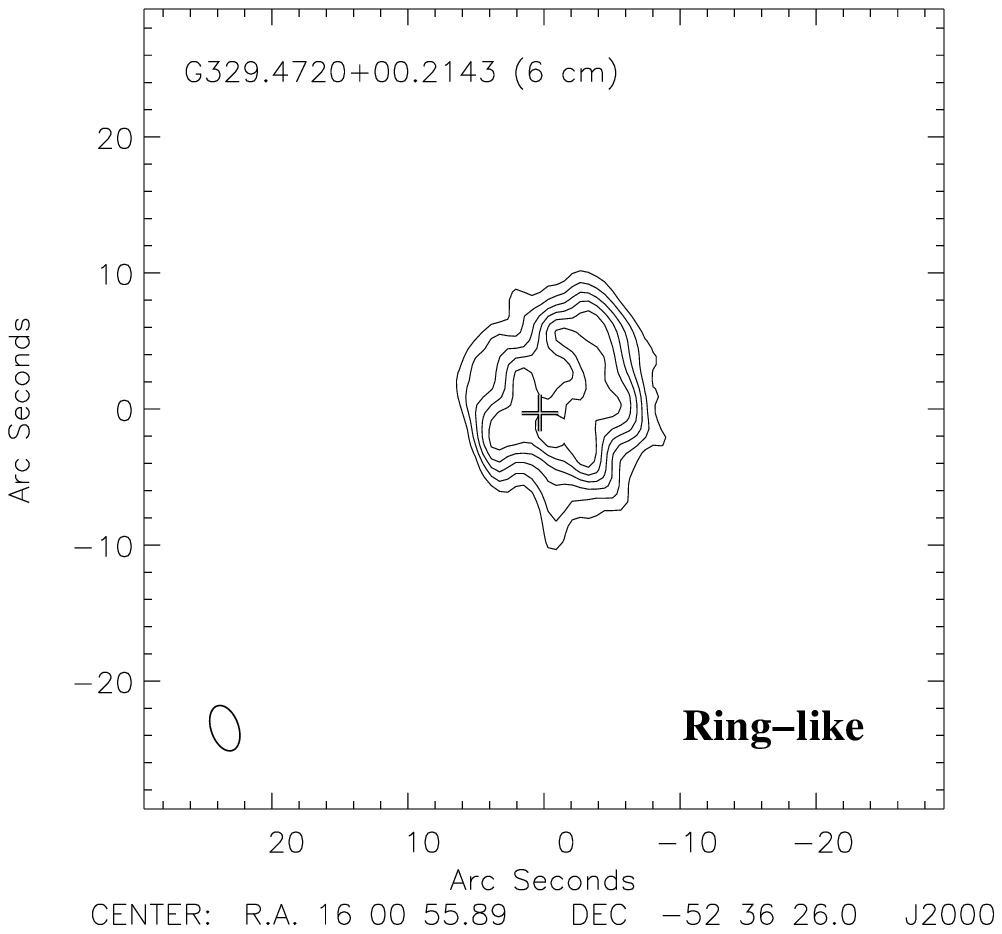}\\ 
\includegraphics[width=0.35\linewidth]{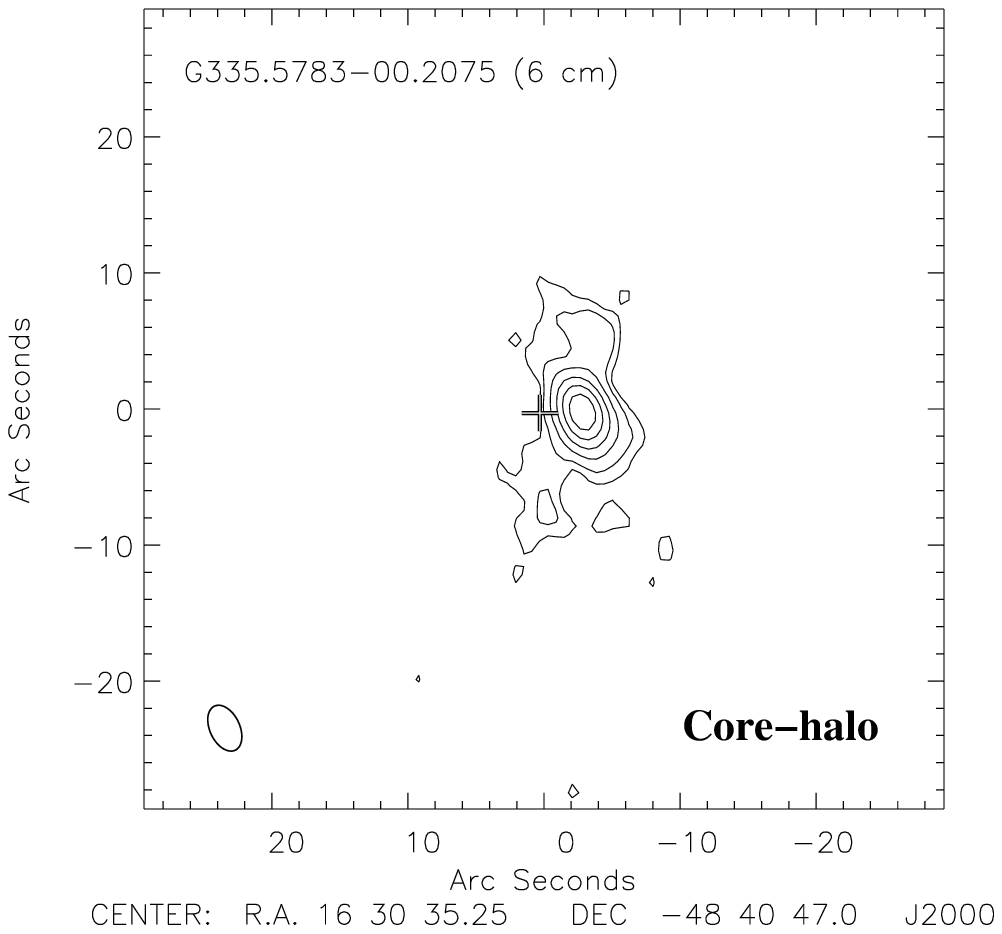} 
\includegraphics[width=0.35\linewidth]{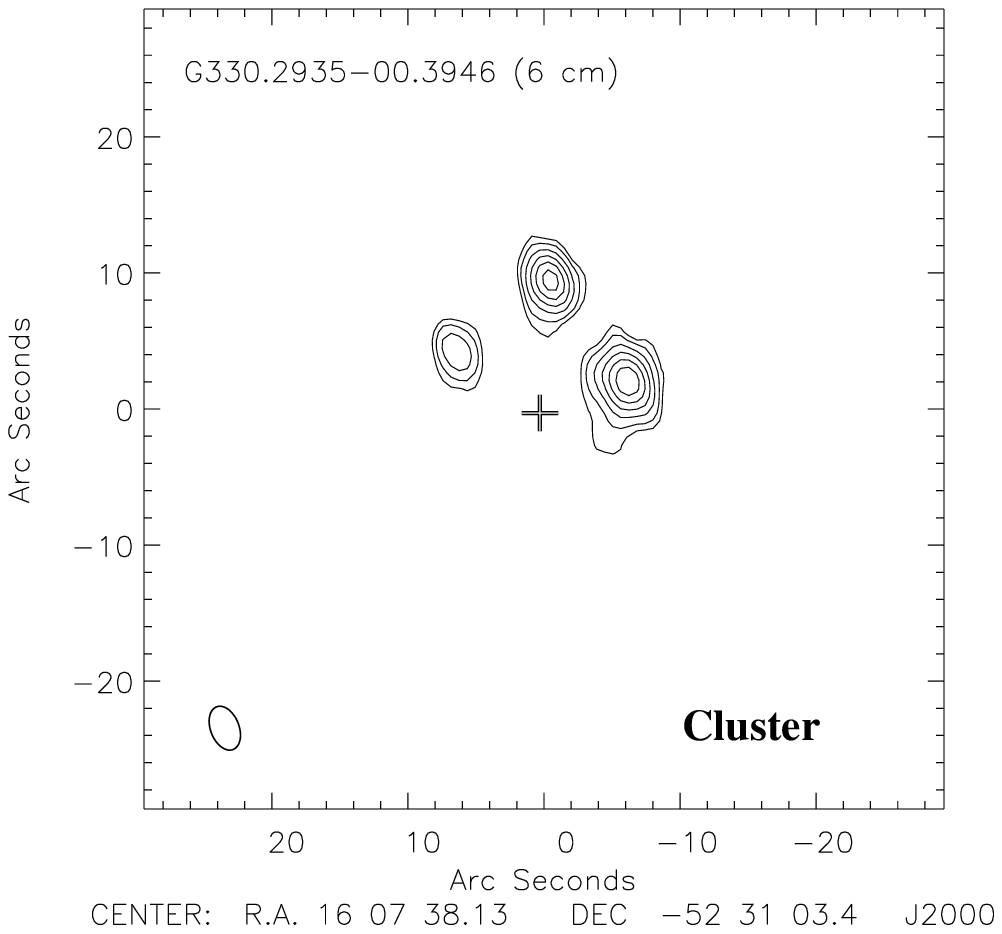}\\ 

\caption{\label{fig:example_maps} Contour radio maps of a sample of the sources detected in this survey. The first contour starts at 4$\sigma$ with the intervening levels determined by a dynamic power law (see text for details). The RMS name and wavelength are given in the top left corner. The position of the RMS point source is indicated by a cross. These sources have been chosen to illustrate the different kinds of source morphologies, as indicated in the lower right corner of each map, and their locations with respect to the nearest RMS point source. The last of these is not a separate morphological type, but is presented here as an example of multiplicity which is discussed in Sect.~\ref{sect:obs_results}.}

\end{center}
\end{figure*}

Each source has been classified by its morphology into one of five types following the classification scheme developed by \citet{wood1989a}, these are: spherical/unresolved (S/U), cometary (C), irregular/multi-peaked (I/MP), shell-like (SH) and core-halo (CH). We note that in a recent deep multi-configuration study of the W49A and Sgr B2 massive star forming regions by \citet{de_pree2005} the core-halo morphological type was abandoned as these sources appeared to be a superposition of a compact source on an unrelated extended source. However, our observations do not have the sensitivity or dynamic range to begin to separate these two components, and to be consistent with previous surveys of this nature (i.e., \citealt{wood1989a,kurtz1994,walsh1998}), the core-halo morphology has been included. The distribution of these morphological types is summarised in Table~\ref{tbl:morphological_classification}. We present a sample of the maps  in  Fig.~\ref{fig:example_maps} to illustrate these different morphological types. Source morphologies are indicated using the symbol in parenthesis in column (3) of Table~5 (in the on line version).
 
\setcounter{table}{3}
\begin{table}[!tbp]
\begin{center}
\caption{Morphological classifications.}
\label{tbl:morphological_classification}
\begin{minipage}{\linewidth}
\begin{tabular}{lcrccr@{.}lc}
\hline
\hline
Morphology & \multicolumn{3}{c}{Number}&		 \multicolumn{4}{c}{Percentage}\\
\hline
Cometary								&	&36 & & & 15&7 	 \\
Core-halo							&	&6  & & & 3&9  	 \\
Shell-like							&	&4  & & & 1&7  	 \\
Irregular/Multi-peaked						&	&29 & & & 13&9 	 \\

Spherical/Unresolved			&	&135&&  &   64&8  	\\
\hline
\end{tabular}
\end{minipage}
\end{center}
\end{table}

At first glance the distribution of morphological types looks different from those presented by \citet{wood1989a} and \citet{kurtz1994}. The first thing to point out is that assigning classifications can be a very subjective exercise as some sources display characteristics of more than one morphological type. A larger number of our sources have unresolved or spherical morphologies but this could either be due to the lower resolution of our observations (i.e., $\sim$1.5--3\arcsec\ compared to $\sim$0.5--0.9\arcsec\ of \citealt{kurtz1994}) or that our selection of unresolved MSX point sources has excluded a larger proportion of the more extended sources. 

If we disregard unresolved and spherical sources, since many of these have been found to possess one of the four more complex  morphologies when imaged at higher resolution (\citealt{wood1989a}), we find the remaining sources have the following distribution: cometary (48\%), irregular/multi-peaked (39\%), core-halo (8\%) and shell-like (5\%). These are in reasonable agreement with previous studies as reviewed by \citet{hoare2005}.

\subsection{Source parameters}
\label{sect:radio_parameters}


In Table~5 we present the observational parameters for each radio source assocaited with an RMS source, i.e., position, peak and integrated fluxes, and sizes of the sources' major and minor axis. Source positions were determined from a two component Gaussian fit to each radio source. In a couple of cases the higher resolution 3.6~cm maps were able to resolve a 6~cm source into two separate components (e.g., RMS source G313.3004+01.1343). For sources classified as having either multi-peaked, irregular or shell-like morphologies the fit has been made to the brightest region of emission, and therefore, the derived position may not necessarily be near the geometric centre of the source. 

Integrated fluxes have been determined by summing the flux within a carefully fitted polygon around each source. As pointed out by \citet{kurtz1994} given the variety of morphologies a single definition of source size is not appropriate. We have therefore followed their procedure for estimating source sizes, i.e., for spherical and unresolved sources the sizes given correspond to the source major and minor axes. For sources classified as cometary the major and minor sizes correspond to the geometric mean diameter at $\sim10\%$ level and the FWHM of a slice through the peak brightness position and perpendicular to the symmetry axis respectively. The major and minor sizes of irregular and multi-peaked sources have been estimated from the approximate angular diameters at the 10--20\% levels of the complex. For shell-like sources the minimum and maximum correspond to the inner and outer diameters of the shell as measured at the half power points and for core-halo sources the major and minor axes of the core have been given.

\subsection{Results of incidental detections}
\label{sect:additional_obs}

A number of observations were made towards MSX sources in the early stages of this project that were later excluded. Althought these data have no direct relevance to the RMS survey we present them here as they may prove useful to the wider astronomical community, however, these results  will not be discussed in detail and for that reason we present them here. 

In total 190 observations were made towards MSX point sources that were later excluded from our sample of MYSOs. Our original colour selection were made using an early version of the MSX point sources catalogue (MSXv1.3), however, when the same selection criteria was applied to the more recent version of the catalogue (MSXv2.3), many of the originally identified sources were found to no longer fulfil our selection criteria and were subsequently dropped. In addition to the sources that failed the colour cuts in the final version of the MSX catalogue there were a number of sources that passed the colour cuts but were found to be associated with 2MASS sources that possess flat or blue near-IR colours, or were found to be considerable extended in the MSX images and therefore more likely to be HII~regions rather than MYSOs.

Since these observations have been made as part of our project, the observational procedures, data reduction and source parameters are all as described in Sections~\ref{sect:rms_observations}, \ref{sect:data_reduction} and \ref{sect:radio_parameters} respectively. The  observation dates and  configuration parameters  are as presented in Tables~\ref{tbl:observation_dates} and \ref{tbl:radio_setup} respectively. The field names, pointing position and parameters derived from the reduced maps are presented in Table~6 (available in electronic form at the CDS via anonymous ftp to cdsarc.u-strasbg.fr (130.79.125.5) or via http://cdsweb.u-strasbg.fr/cgi-bin/qcat?J/A+A/). 

\begin{figure}
\begin{center}
\includegraphics[width=0.45\textwidth]{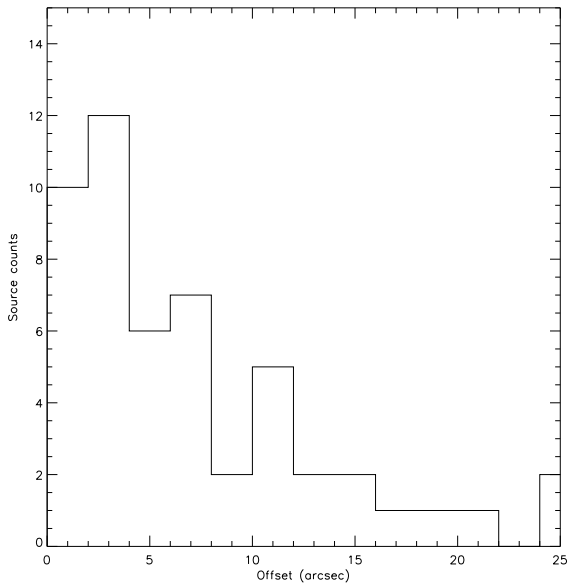}
\caption{Histogram showing the projected angular separations between the incidental radio detections (see text for details) and their nearest MSX point source counterpart. The positional offsets for all possible matches within the 25\arcsec\ are plotted.}

\label{fig:msx_radio_offsets}
\end{center}
\end{figure}

In total an additional 169 radio sources were detected, 51 of these were found in RMS fields and the 118 were found in fields towards later discarded MSX sources; as these detections are not related to our survey we consider these to be incidental detections. We searched the MSX point source catalogue out to a radius of 25\arcsec\ for possible mid-IR counterparts and used the same 12\arcsec\ cutoff used to identify genuine associations from chance alignments (see Sect.~\ref{sect:obs_results}). In Fig.~\ref{fig:msx_radio_offsets} we present a plot of the angular separation between the radio sources and their nearest MSX counterpart. 

\begin{figure*}
\begin{center}

\includegraphics[width=0.35\linewidth]{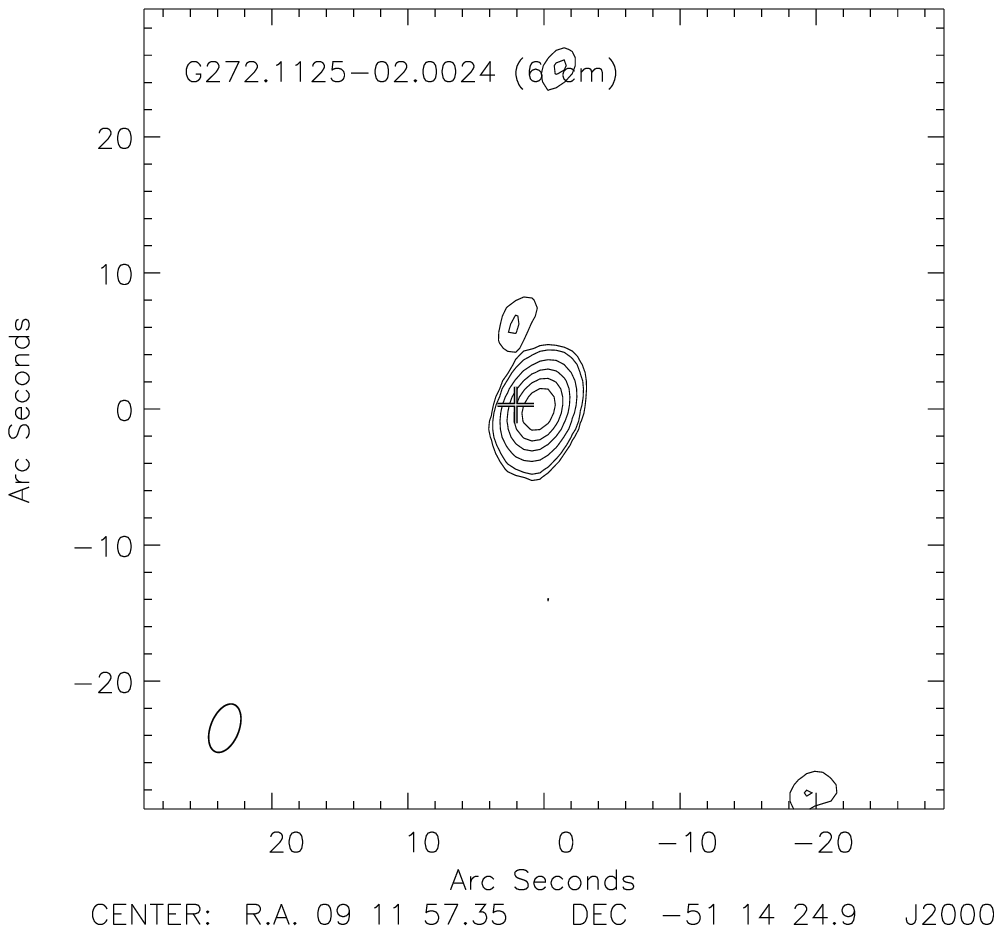} 
\includegraphics[width=0.35\linewidth]{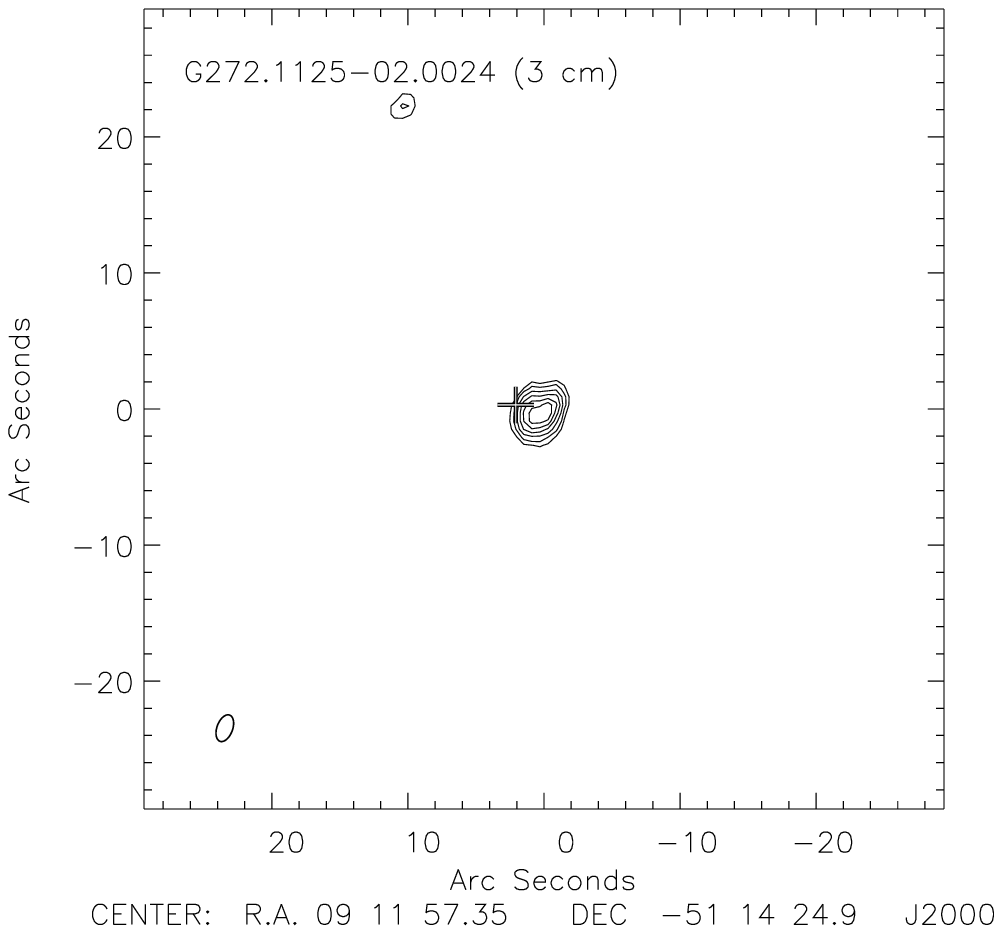}\\ 
\includegraphics[width=0.35\linewidth]{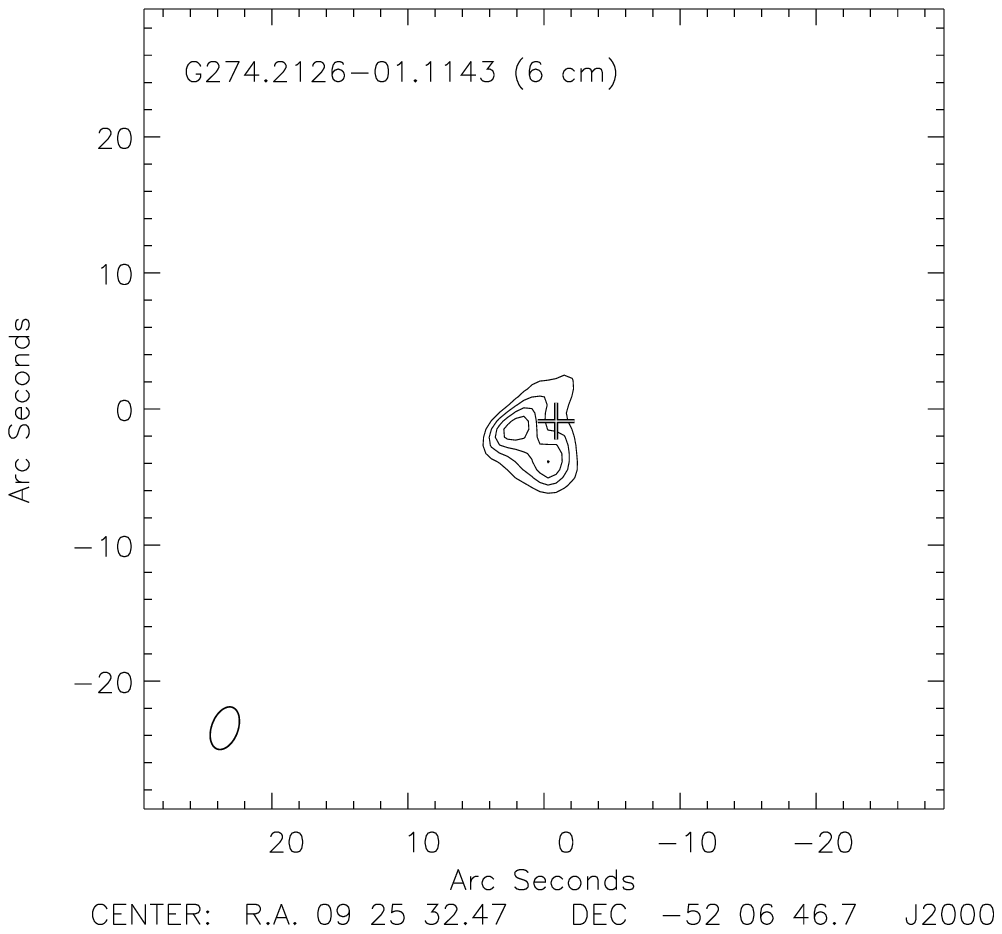} 

\includegraphics[width=0.35\linewidth]{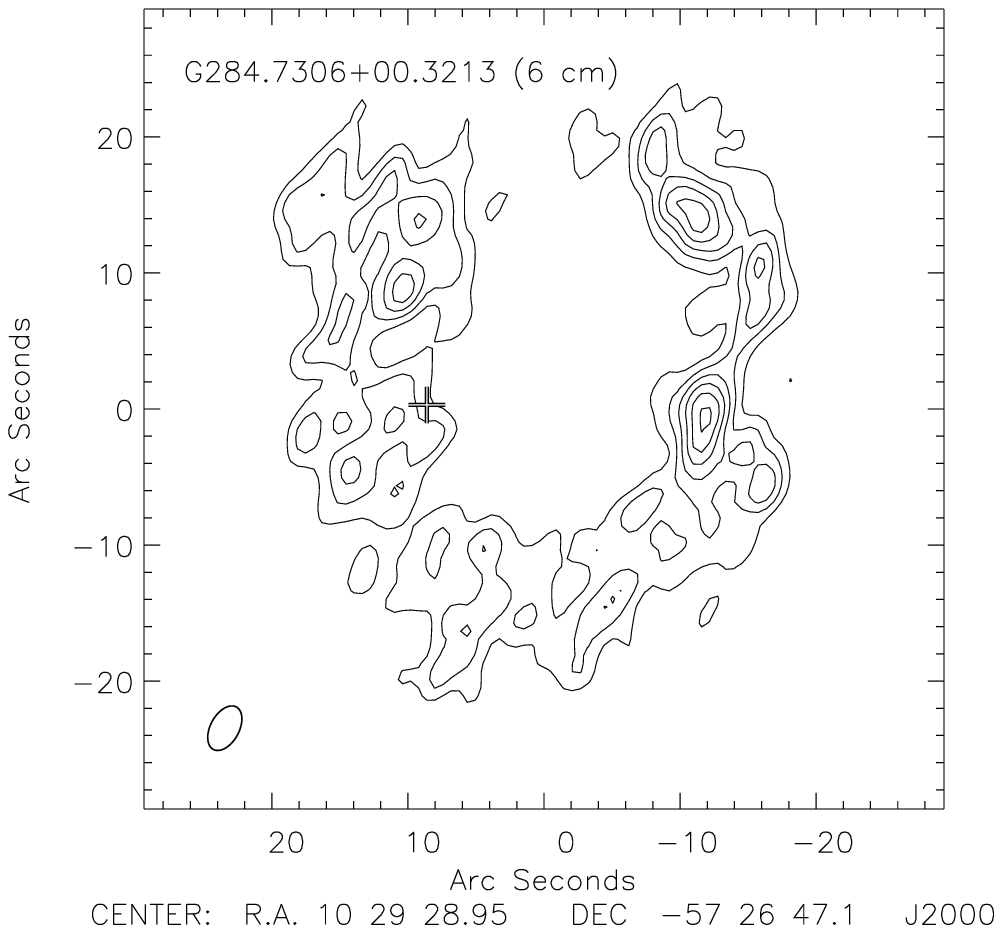}\\ 
\caption{\label{fig:msx_maps} Contour radio maps of the radio sources with MSX counterparts. The first contour starts at 4$\sigma$ with the intervening levels determined by a dynamic power law (see text for details). The field name and wavelength are given in the top left corner. The size of the synthesised beam is shown to scale in the lower left hand corner. The full version of this figure is only available in electronic form at the CDS via anonymous ftp to cdsarc.u-strasbg.fr (130.79.125.5) or via http://cdsweb.u-strasbg.fr/cgi-bin/qcat?J/A+A/.}

\end{center}
\end{figure*}

Although this plot has been produced with a relatively small sample it shares the same characteristics as the distribution of separations of RMS-radio matched sources presented in Fig.~\ref{fig:rms_radio_offsets}, i.e., strongly peaking at $\sim$2\arcsec\ and tailing off $\sim$12\arcsec. From this plot we see that 12\arcsec\ is a reasonable cutoff as it was for the RMS-radio matched sources discussed earlier. If we consider all matches within 12\arcsec\ as genuine matches, as we did for the RMS-radio associations, we find radio emission associated with 37 MSX sources, 5 of which are associated with two radio sources. 42 Radio sources were matched in total. The majority of these MSX-radio matched sources passed the initial colour selection, and therefore have mid-IR colours consistent with being either UCHII regions or potential MYSO candidates, but were subsequently found to be extended rather than point sources in the MSX images, or failed the near-IR colour selection and are therefore more likely to be UCHII regions. The derived source parameters for each radio source found to have an MSX counterpart are presented in Table~7 (in the on line version) and contour plots of the emission are presented in Fig.~\ref{fig:msx_maps}. These plots are as described in Sect.~\ref{sect:emission_maps} with the MSX source name given in the upper left corner. These matched radio-MSX sources are almost certainly Galactic and as with the radio-RMS sources are most likely to be young compact HII regions.

\begin{figure*}
\begin{center}

\includegraphics[width=0.35\linewidth]{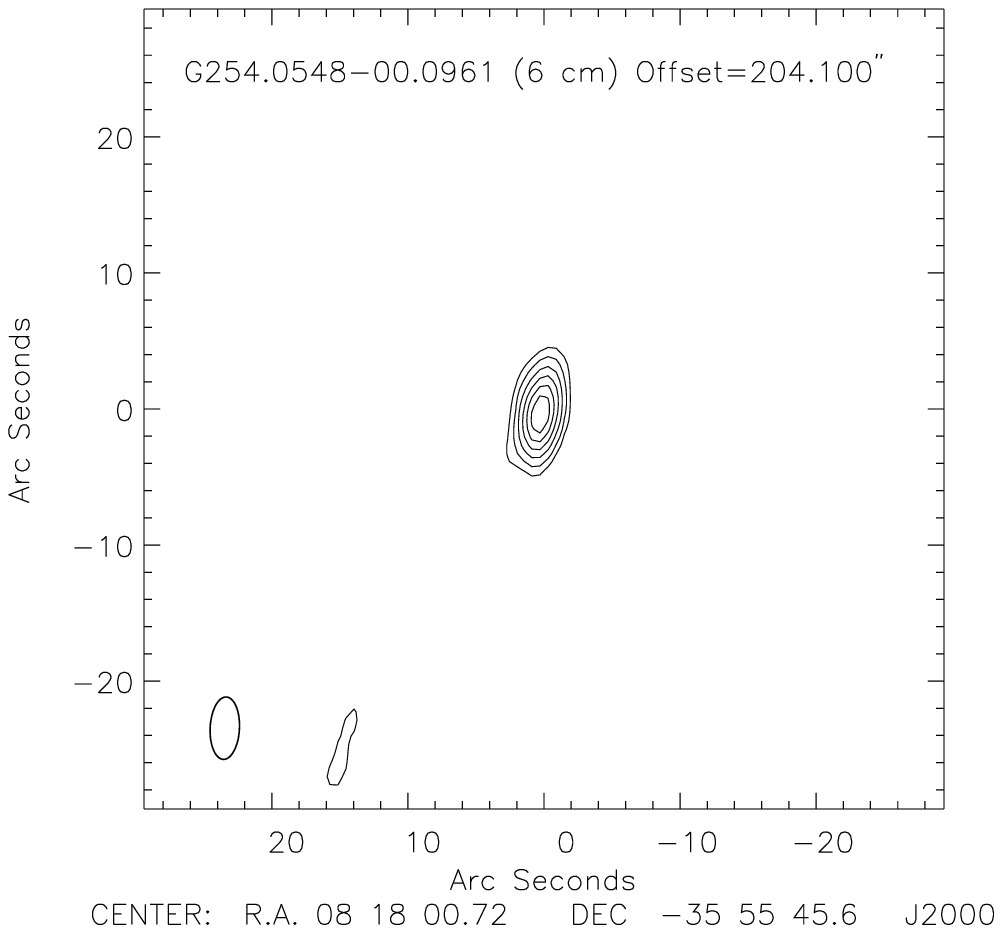} \\
\includegraphics[width=0.35\linewidth]{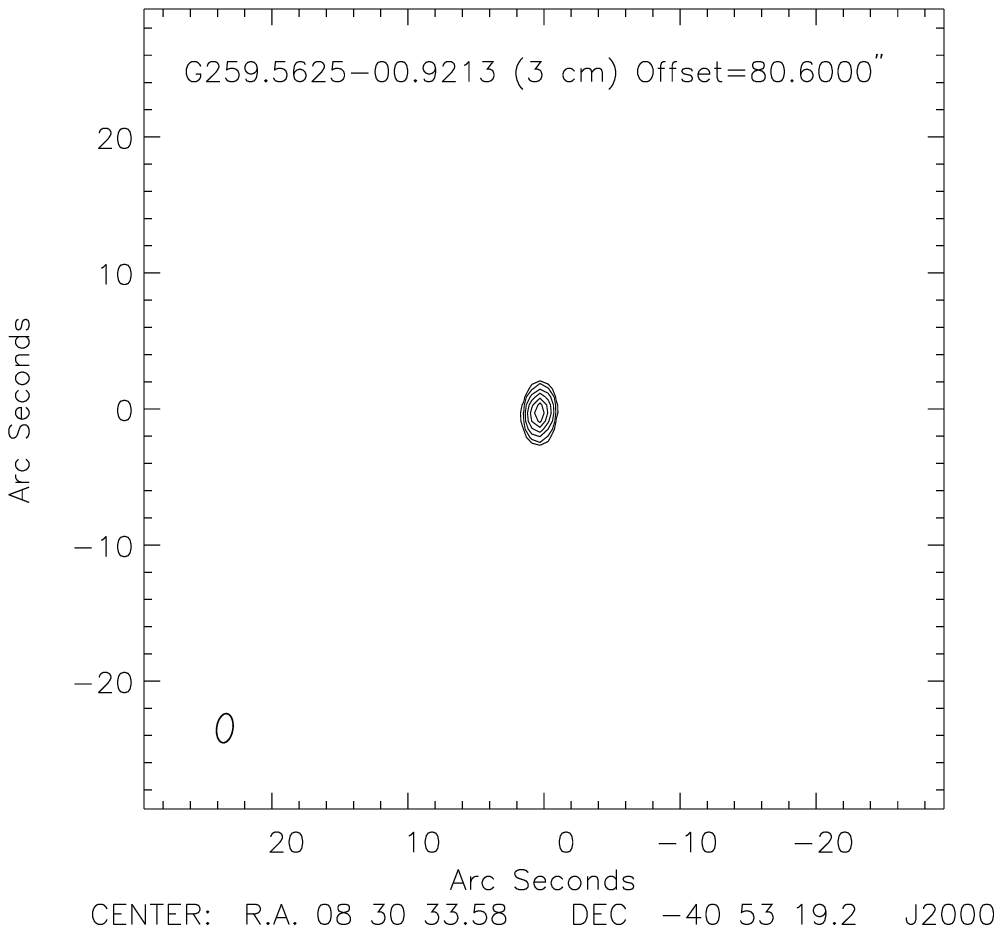}\\ 
\includegraphics[width=0.35\linewidth]{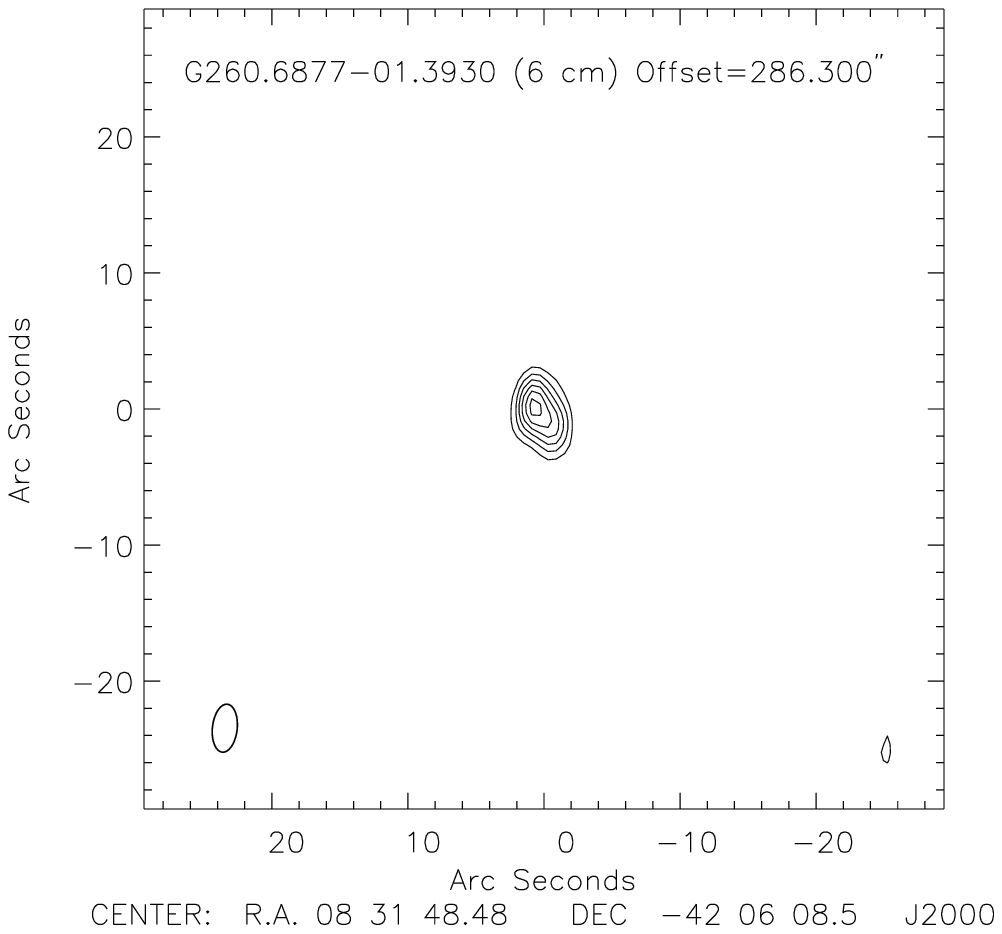}\\ 

\caption{\label{fig:unmatched_maps} Contour radio maps of the radio sources with no MSX counterparts. The first contour starts at 4$\sigma$ with the intervening levels determined by a dynamic power law (see text for details). The MSX name and wavelength are given in the top left corner along with the offset from the centre of the field. The size of the synthesised beam is shown to scale in the lower left hand corner. The full version of this figure is only available in electronic form at the CDS via anonymous ftp to cdsarc.u-strasbg.fr (130.79.125.5) or via http://cdsweb.u-strasbg.fr/cgi-bin/qcat?J/A+A/.}

\end{center}
\end{figure*}

The nature of the remaining 127 radio sources we were unable to identify a mid-IR counterpart is uncertain as there are several kinds of object that could be responsible for the detected emission e.g., radio stars, weak mid-IR sources or extragalactic background sources. Most known radio stars have fluxes below the sensitivity of our observations and so these are not considered to make a significant contribution. We would suggest that the majority of these sources are extragalactic background objects.  The derived source parameters for these unmatched radio sources are presented in Table~8 (in the on line version) and contour plots of the emission are presented in Fig.~\ref{fig:unmatched_maps}. These plots are as described in Sect.~\ref{sect:emission_maps} except the observational field name and the sources offset from the observation centre are given in the upper left corner.

\section{Discussion}
\label{sect:discussion}
\subsection{Distribution of radio sources}

\begin{figure*}
\begin{center}
\includegraphics[width=0.95\linewidth]{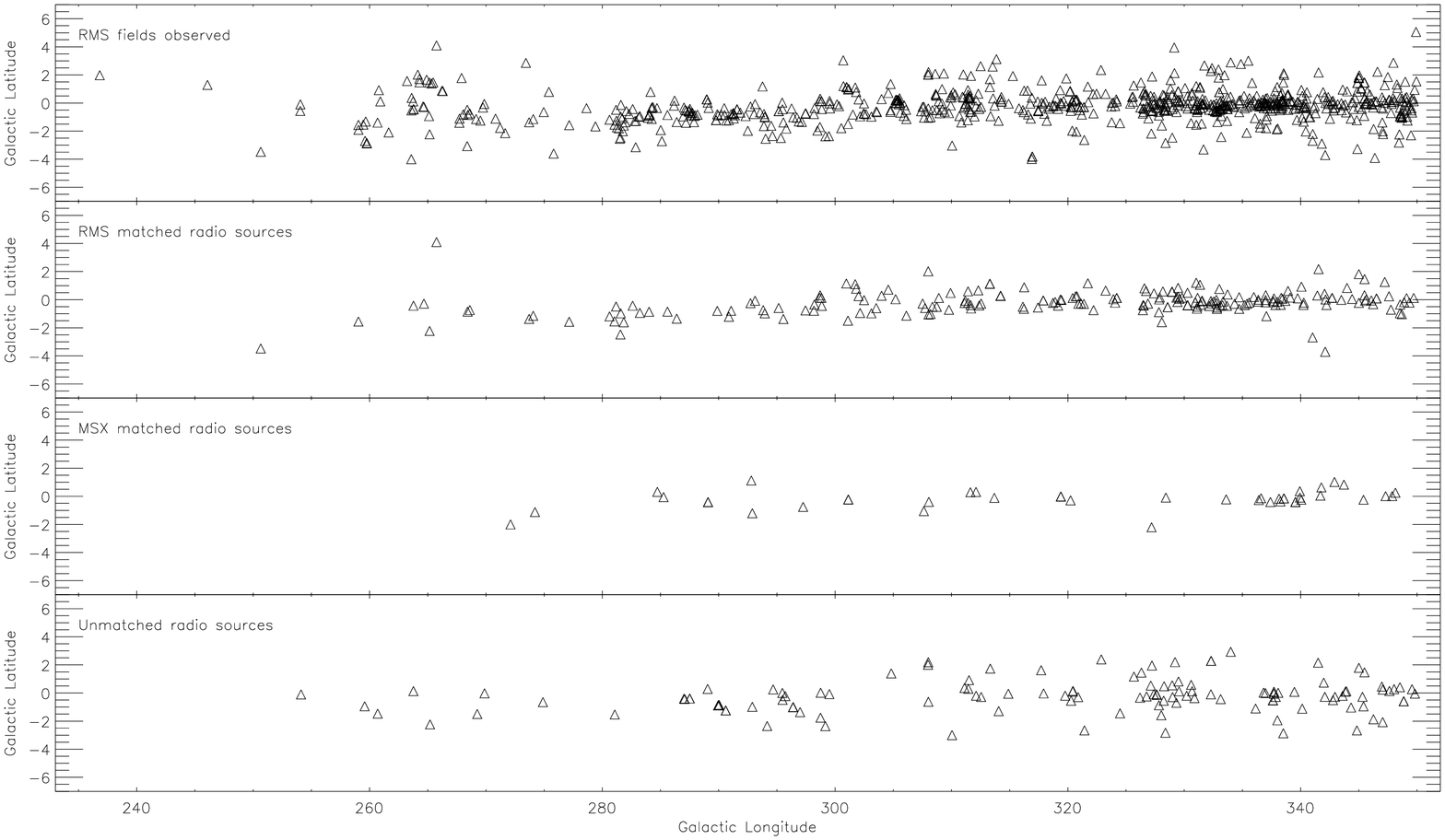}
\caption{\label{fig:galactic_distirbution} Galactic distribution of the 826 RMS sources observed with ATCA (upper panel), RMS sources associated with radio emission (upper middle panel),  serendipitously matched MSX and radio sources (lower middle panel) and unmatched radio sources (lower panel).}

\end{center}
\end{figure*}

In the \emph{upper panel} of Fig.~\ref{fig:galactic_distirbution} we present a plot of the two dimensional distribution of all RMS sources observed with the ATCA, followed in the \emph{upper middle}, \emph{lower middle} and \emph{lower panels} by the RMS sources associated with radio emission, serendipitously matched MSX and radio sources, and the unmatched radio sources. This figure shows the distribution of the RMS sources to be correlated with the Galactic plane. However, the RMS sources that are associated with radio emission display an even tighter correlation with the Galactic plane. As previously mentioned, the vast majority of RMS sources associated with radio emission are likely to be UCHII regions and the tight correlation with the Galactic plane supports this. The serendipitously matched MSX and radio sources are also found to be located within a very limited range of Galactic latitudes which would suggest that most of these are also UCHII regions. The distribution of the unmatched radio sources shows no such correlation with the Galactic plane and are fairly evenly spread as would be expected if these sources are extragalactic background sources as suggested in the previous section.



\begin{figure}
\begin{center}
\includegraphics[width=0.45\textwidth]{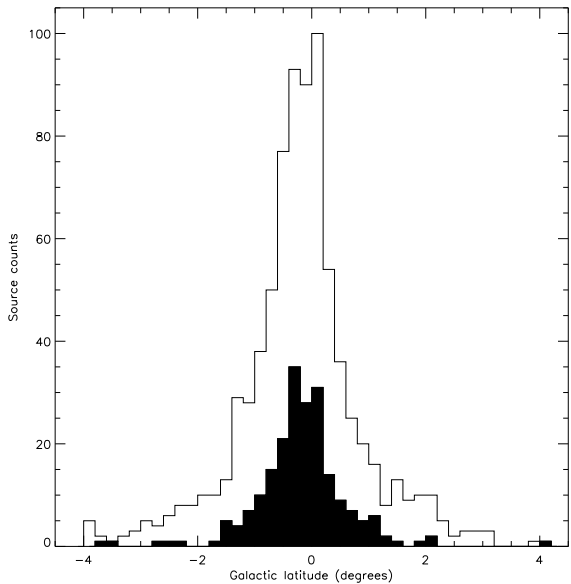}
\caption{\label{fig:latitude_distribution} Histogram of the latitude distribution of the MYSO candidates observed in this survey. We have over-plotted the latitude distribution of the MYSO candidates with associated radio emission as a filled histogram.}

\end{center}
\end{figure}

The correlation of RMS-radio associated sources with the Galactic plane is better illustrated in Fig.~\ref{fig:latitude_distribution}. In this figure we present a histogram of the number of RMS sources observed as a function of Galactic latitude. We have over-plotted the RMS sources with associated radio emission (filled histogram). In principle we should expect the distribution of RMS sources associated with radio emission to mimic the distribution of the RMS sample as a whole, and in general this is the case. However, although the distributions look similar the RMS-radio matches have a smaller latitude spread. The angular scaleheight of the southern RMS catalogue is $\sim$0\degr.7 whereas the scaleheight of the RMS UCHII regions is slightly smaller at $\sim$0\degr.6. The larger scaleheight for the sample as a whole is probably the result of a residual contamination by dusty evolved stars at high latitudes and from nearby low-mass YSOs as predicted by \citet{lumsden2002}. Our molecular line and near-IR spectroscopy observations will provide a means for assessing the nature of these high latitude RMS sources.

The angular scaleheight of the UCHII regions is in excellent agreement with the value found by \citet{wood1989a} who reported a scaleheight of 0.6\degr\ from IRAS identification of $\sim$1600 potential UCHII regions. At the Galactic centre distance this corresponds to a scaleheight of 89 pc. This value is significantly broader than the scaleheight of $\sim$0\degr.25 found from the 1.4 and 5 GHz VLA Galactic plane surveys reported by \citet{giveon2005b} and \citet{giveon2005a} respectively. However, these surveys were restricted to small latitude ranges ($|b| \le$1\degr.8 and $|b| \le$ 0\degr.4 for the 1.4 and 5 GHz surveys respectively), which could explain their smaller derived scaleheight.                                                           


\subsection{Comparison of source fluxes with other surveys}
\label{sect:integrity_check}

Although most sources that had been previously observed at high resolution were removed, a small number were left in as an integrity check on our results. Ten of the sources we detected had previously been detected in a radio continuum survey of methanol maser sources reported by  \citet{walsh1998}. In Table~\ref{tbl:integrity_check} we present a comparison of the 8.6 GHz fluxes reported in this paper and those reported for the same sources by \citet{walsh1998}.

\renewcommand{\thefootnote}{\alph{footnote}}
\setcounter{footnote}{0}
\setcounter{table}{8}
\begin{table}
\begin{minipage}[t]{\linewidth}

\begin{center}

\caption{Comparison of sources detected at 8.6~GHz in this survey and in the methanol maser survey of \citet{walsh1998}. (The peak fluxes for the methanol maser survey have been taken from \citet{walsh1998}, however, the integrated fluxes were not presented in their paper but are available from VizieR (http://vizier.u-strasbg.fr/viz-bin/VizieR).)}
\label{tbl:integrity_check}
\begin{tabular}{lrrrr}
\hline
\hline
  &\multicolumn{2}{c}{RMS survey}& \multicolumn{2}{c}{\citet{walsh1998}} \\
  RMS name& Peak&	Integ. & Peak&	Integ.\\ 

\hline 
G268.4222$-$0.8490&307.1&1,696.0&510 &1,012.0 \\
G305.1967+0.0335&113.1&1,105.0&90& 394.0\\
G318.9148$-$0.1647&60.6&1,659.0&120 &424.0 \\
G330.9544$-$0.1817&591.8&1,728.0&280&1,161.0 \\
G337.7051$-$0.0575&152.8&171.0&84 & 94.0 \\
G344.4257+0.0451&63.5&268.3&28& 485.0\\
G345.0061+1.7944&189.9&209.0&140& 130.0 \\
G345.4881+0.3148&456.9&2,163.0&200& 2,100$^a$ \\
G346.5234+0.0839&77.0&150.3&45& 104.0 \\
G348.6972$-$1.0263&402.2&1,194.0&500& 843.0 \\
\hline
\end{tabular}
\end{center}
$^a$ The integrate flux for this source obtained from VizieR is incorrect (48.4 mJy), the correct flux was provided by Andrew Walsh (private com.).
\end{minipage}
\end{table}

We expected to find variations of $\sim$10--20\%, mainly due to calibration uncertainties, however, looking at Table~\ref{tbl:integrity_check} it is immediately apparent that the differences are much larger than expected with our fluxes being, in general, approximately twice those reported by \citet{walsh1998}, and in a few cases three or four times larger. The main reason for these different integrated flux measurements is that different array configurations are sensitive to different angular scales, with compact arrays being more sensitive to extended emission, and more extended arrays filtering out the larger scale structure and being more sensitive to compact high brightest features. Although both surveys used a 6 km array configuration the baseline lengths were quite different; \citet{walsh1998} used the 6A configuration which has a minimum baseline of 337 m, which is between two and four times longer than the minimum baselines of the  6C and 6D arrays used for our observations respectively. The arrays used for our observations are therefore more sensitive to extended structures than the array used by \citet{walsh1998} and have resulted in higher integrated fluxes being measured. 

This point can be illustrated using the results reported by \citet{ellingsen2005} for G318.9148$-$0.1647, a source also common to both our observations and those of \citet{walsh1998}. \citet{ellingsen2005} observed a number of sources in two different configurations with the ATCA designed to investigate the ratio of extended to compact emission associated with UCHII regions. They used the 750D and the 6A configurations which have minimum and maximum baselines of 31 m + 719 m, and 337 m + 6 km respectively. For G318.9148$-$0.1647 \citet{ellingsen2005} report an integrated flux of 1078 mJy and 1513 mJy for the 6A and 750D arrays respectively. The shortest baseline on the 6D array used for our observations of this particular source is comparable to that of the 750D array used by \citet{ellingsen2005}, which explains why similar integrated fluxes were obtained. It is not clear why the integrated fluxes reported by \citet{ellingsen2005} for their 6 km observations and by \citet{walsh1998} are so different (1078 mJy and 428 mJy respectively) since the observations were carried out using the same configuration. However, we note that the observations made by \citet{ellingsen2005} had six times more on-source integration time than ours or those of \citet{walsh1998}. This analysis highlights that fluxes measured by an interferometer are extremely sensitive to both the amount of on-source integration time and the array configurations used.

\subsection{Radio spectral indices}

\begin{figure}
\begin{center}
\includegraphics[width=0.45\textwidth]{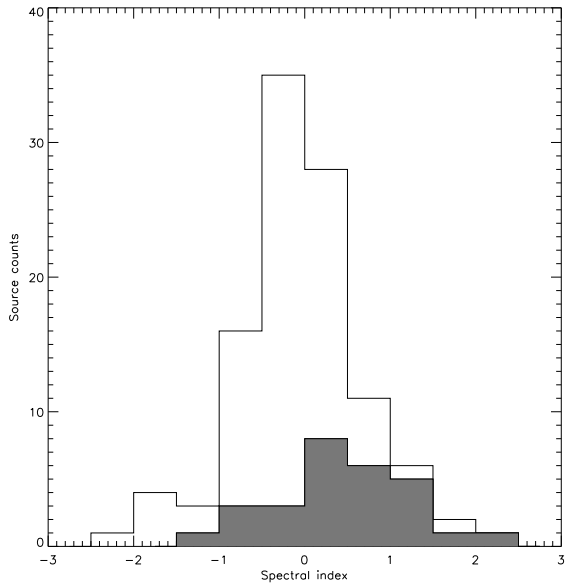}
\caption{\label{fig:spectral_index} Plot of the spectral indices RMS-radio sources detected at both frequencies (unfilled histogram) over-plotted with the spectral indices calculated for unresolved radio sources only (grey histogram). (Source classified as having an irregular morphology have been excluded.) Note both distributions are slightly skewed towards positive spectral indices, suggesting the detection of a significant number of nebulae that are optically thick to radio emission.}
\end{center}
\end{figure}

In Fig.~\ref{fig:spectral_index} we present a plot of the spectral indices (F$_\nu \propto \nu^\alpha$) for all RMS-radio sources detected at both frequencies that have not been classified as having an irregular morphology. The complete sample is shown as an unfilled histogram. We note that the ATCA is not a scaled array between the two frequencies observed and therefore each frequency is sensitive to different spatial scales. This rendered the spectral indices for individual resolved sources unreliable, especially for some of the more complex morphologies (i.e., cometary and core-halo morphologies). However, the combined spectral indices should reflect the statistical behaviour of the sample relatively accurately and indicate trends in the data. 

There are three main features that are apparent in this figure. Firstly, the  spectral indices peak between -0.5 and 0 and have an average  of -0.16, consistent with the emission detected from thermal sources (e.g., UCHII regions). Secondly, the distribution is slightly skewed in the direction of positive spectral indices, possibly indicating a significant number of optically thick sources have been detected. Thirdly, it is surprising to find a large number of RMS sources associated with radio sources that display large negative spectral indices. 


In order to check these trends we have over-plotted the spectral indices  of sources that are unresolved at both frequencies (grey histogram) on Fig.~\ref{fig:spectral_index}. Even though the number counts for this histogram are limited we can see that the main trends seen in the distribution for the whole sample are still evident, however, the proportion of RMS-radio matched sources with positive indices has increased significantly and the peak of the distribution has shifted from $\sim$ $-$0.16 to +0.16. There are still a number of RMS matches that are associated with radio sources that have steep inverted spectral indices, but again the proportion, and the magnitude of the negativity has significantly decreased compared to the sample as a whole. It is possible that some of these RMS-radio matches are chance alignments with background radio source.

\subsection{Single frequency detections}
\label{sect:single_frequency_detections}

Approximately one third of the radio sources associated with RMS sources were only detected at one of the two observed frequencies. The majority of these, 45, were detected at 6~cm with a further seven detected at 3.6~cm. There are a number of possible explanations for a sources to be detected at one frequency and not the other. For example, extragalactic background sources and supernova remnants have steep inverted spectral indices and are therefore less likely to be detected at the high frequency. Alternatively, if the emission is from an optically thick nebula, such as an UCHII region, it is more likely to be detected at the higher frequency. Therefore further investigation of the reasons behind these single frequency detections could lead to new insight into the nature of these sources; some of these possibilities are discussed below.

\subsubsection{6~cm detections}

In several cases, when observing two or more RMS sources located in the same field of view, a source was located outside the primary beam of the 3~cm observation and was therefore only detected at 6~cm. Eleven 6~cm detections can be explained this way. Another consideration is the interferometer's sensitivity to angular scales, which as previously mentioned is different for the two frequencies used for these observations. As previously mentioned the largest angular size that can be accurately imaged at 3.6 and 6~cm is $\sim$20\arcsec\ and $\sim$30\arcsec\ respectively. It is therefore possible for an extended source to be successfully imaged at 6~cm, but either poorly imaged or totally resolved out at 3.6~cm. Examination of the maps presented in Fig.~3 reveals that 26 of these single 6~cm detections are larger than the maximum imagible at 3.6~cm (e.g., G281.5857-00.9706, G338.2900-00.3729 and G343.5024-00.0145), and were resolved out in the 3.6~cm maps. 

In total 37 sources detected only at 6~cm can be explained by either the difference in the size of the 3.6 and 6~cm field of views or  sources being too extended to have been accurately imaged at 3.6~cm. The remaining eight sources would need to have  steep inverted spectral indices in order for them not to have been detected in the 3.6~cm maps, which would suggest these radio source to be extragalactic in origin. For these sources we find typical positional separations between the RMS and radio sources of $\sim$2\arcsec, and are therefore coincident within the errors -- better than the sample in general. From our analysis of the correlation between the radio and RMS sources we determined the maximum expected number of false matches to be four. The spectral index analysis presented in the previous subsection also revealed a number of radio sources with negative spectral indices associated with RMS sources. It is possible that some of these are chance alignments of radio sources with our RMS sources, however, the excellent positional correlation and low probability of chance associations cannot explain them all.

\subsubsection{3~cm detections}

Seven sources were detected at 3.6~cm, all of which have fluxes significantly larger than the corresponding 4$\sigma$ rms noise level at the same position in their corresponding 6~cm maps. Using these noise levels we calculated a lower limit for the spectral index ($F_\nu\propto\nu^\alpha$) for each of these sources and found them all to be $\alpha \ge 1$ with the majority lying between 1 and 2, consistent with these objects being optically thick. The presence of a number of optically thick sources was also indicated by the  histogram of spectral indices presented in the previous subsection. The circumstantial evidence points to these sources being deeply embedded UCHII regions, possibly at a relatively early stage in their evolution.

\section{Summary and conclusions}
\label{sect:summary}

These observations form part of a multi-wavelength programme of follow-up observations of a sample of $\sim$2000 colour selected MYSO candidates designed to distinguish between genuine MYSOs and other embedded or dusty objects. In this paper we report the results of radio continuum observations towards a sample of MYSO candidates located in the southern sky (235\degr $<$ $l$ $<$ 350\degr). Observations were made at 3.6 and 6~cm towards 826 RMS sources located within 802 fields using the ATCA. These observations were aimed at identifying radio loud contaminants such as UCHII regions and PNe from the relatively radio quiet MYSOs. These observations had typical rms noise values of $\sim$0.3~mJy and were therefore sensitive enough to detect a HII region powered by B0.5 or earlier type star located at the far side of the Galaxy. Our main finding are as follows:

\begin{enumerate}

\item  Of the 826 RMS sources observed we have found 199 to be associated with radio emission. In total we have found that contamination of our MYSO sample by radio loud sources is $\sim$25\%. More interestingly we failed to detect any radio emission towards 627 RMS sources and therefore after eliminating one of the main sources of contamination we are still left with a large sample of MYSOs candidates.

\item Our colour selection criteria allowed us to identify a large proportion of PNe with similar MSX colours but were not so effective at identifying contamination by UCHII regions. Therefore the majority of the 199 RMS sources found to be associated with radio emission are expected to be UCHII regions. The morphologies of these sources, which are consistent with other studies of UCHII regions, relatively flat spectral indices, and their scaleheight certainly support their identification as UCHII regions.

\item In addition to the 211 radio sources detected towards RMS sources we detected a further 169 discrete radio sources. 51 of these were found within RMS fields and the remaining 118 were found within fields that were observed as part of our programme, but no longer fulfil our selection criteria. Cross-correlating the positions of these field detections with the MSX point source catalogue identified 42 sources to be associated with 37 MSX sources. These radio-MSX matched sources are distributed over a small range in Galactic latitudes, similar to the RMS-matched sources, and are therefore predominately thought to be UCHII regions. The remaining 127 radio sources for which no mid-IR source could be identified are thought to consist mainly of extragalactic background objects.

\end{enumerate}

\begin{acknowledgements}

The authors would like to thank the Director and staff of the Paul Wild Observatory, Narrabri, New South 
Wales, Australia for their hospitality and assistance during the Compact Array. 
JSU is supported by a PPARC postdoctoral fellowship grant. We would also like to thank the referee Michael Burton for his comments and suggestions. This research would not have 
been possible without the SIMBAD astronomical database service operated at CDS, Strasbourg, France and
the  NASA Astrophysics Data System Bibliographic Services. This research makes use of data products from the Two Micron All Sky Survey, which is a joint project of the University of Massachusetts and the Infrared Processing and Analysis 
Center/California Institute of Technology, funded by the National Aeronautics and Space Administration and
the National Science Foundation. This research also made use of data products from the Midcourse Space 
Experiment.  Processing of the data was funded by the Ballistic Missile Defense Organization with additional 
support from NASA Office of Space Science.

\end{acknowledgements}


\bibliography{rms_atca_radio_new_v2}

\bibliographystyle{aa}
\Online

\clearpage
\onecolumn
\setcounter{table}{4}
\include{radio_table6_rms_matches_new}

\setcounter{table}{6}
\include{radio_table7_msx_matches_new}

\include{radio_table8_unmatched}

\end{document}

%% file: radio_table6_rms_matches_new.tex
\begin{longtable}{lccccccrrrrr}
\caption{\label{tbl:radio_parameters}    Observational parameters for each RMS source associated with radio emission.} \\ 
\hline
\hline
 & && & &\multicolumn{2}{c}{Source position}& Peak&	Integ. &\multicolumn{2}{c}{Source} \\ 
RMS  name&&Offset &Morph.&$\lambda$&$\alpha$ (J2000)  &$\delta$ (J2000)&S$_\nu$&S$_\nu$ & \multicolumn{2}{c}{Diameter } \\ 
&&(\arcsec)&&(cm)&(h m s)&($\degr$~\arcmin\ \arcsec\ ) & (mJy/b)&(mJy)& Max (\arcsec) &Min (\arcsec) \\ 
\hline
\endfirsthead
\caption{continued.} \\ 
\hline
\hline
 & && & &\multicolumn{2}{c}{Source position}& Peak&	Integ. &\multicolumn{2}{c}{Source} \\ 
RMS name&&Offset &Morph.&$\lambda$&$\alpha$ (J2000)  &$\delta$ (J2000)&S$_\nu$&S$_\nu$ & \multicolumn{2}{c}{Diameter } \\ 
&&(\arcsec)&&(cm)&(h m s)&($\degr$~\arcmin\ \arcsec\ ) & (mJy/b)&(mJy)& Max (\arcsec) &Min (\arcsec) \\ 
\hline
\endhead
\hline
\endfoot
G250.6510$-$03.4744 && 5.2 & U& 3.6& 07:54:56.11 & $-$34:49:37.9 & 6.7 & 14.3 & 3.3 & 1.9 \\ 
 &&  & & 6.0& 07:54:56.10 & $-$34:49:38.1 & 13.6 & 23.3 & 5.2 & 2.5 \\ 
G259.0453$-$01.5559 && 10.6 & I/MP& 6.0& 08:26:17.54 & $-$40:48:35.0 & 2.5 & 21.7 & 8.7 & 3.6 \\ 
G263.7759$-$00.4281 && 1.9 & U& 6.0& 08:46:34.85 & $-$43:54:29.8 & 1.6 & 1.5 & 5.8 & 1.8 \\ 
G264.6712$-$00.2839 && 3.5 & C& 3.6& 08:50:21.37 & $-$44:30:43.8 & 3.8 & 8.4 & 3.4 & 1.5 \\ 
 &&  & & 6.0& 08:50:21.39 & $-$44:30:41.2 & 13.4 & 36.0 & 6.0 & 3.0 \\ 
G265.1591$-$02.2284 && 0.9 & U& 3.6& 08:43:28.08 & $-$46:06:39.8 & 78.9 & 89.7 & 2.8 & 1.1 \\ 
 &&  & & 6.0& 08:43:28.07 & $-$46:06:39.8 & 75.6 & 83.2 & 4.5 & 1.5 \\
\end{longtable}

%% file: radio_table7_msx_matches_new.tex
\begin{longtable}{lccccccrrrrr}
\caption{\label{tbl:msx_radio_parameters}   Observational parameters for each MSX source associated with radio emission.} \\ 
\hline
\hline
 & && & &\multicolumn{2}{c}{Source position}& Peak&	Integ. &\multicolumn{2}{c}{Source} \\ 
MSX  name&&Offset &Morph.&$\lambda$&$\alpha$ (J2000)  &$\delta$ (J2000)&S$_\nu$&S$_\nu$ & \multicolumn{2}{c}{Diameter } \\ 
&&(\arcsec)&&(cm)&(h m s)&($\degr$~\arcmin\ \arcsec\ ) & (mJy/b)&(mJy)& Max (\arcsec) &Min (\arcsec) \\ 
\hline
\endfirsthead
\caption{continued.} \\ 
\hline
\hline
 & && & &\multicolumn{2}{c}{Source position}& Peak&	Integ. &\multicolumn{2}{c}{Source} \\ 
MSX name&&Offset &Morph.&$\lambda$&$\alpha$ (J2000)  &$\delta$ (J2000)&S$_\nu$&S$_\nu$ & \multicolumn{2}{c}{Diameter } \\ 
&&(\arcsec)&&(cm)&(h m s)&($\degr$~\arcmin\ \arcsec\ ) & (mJy/b)&(mJy)& Max (\arcsec) &Min (\arcsec) \\ 
\hline
\endhead
\hline
\endfoot
G272.1125$-$02.0024 && 1.3 & U& 3.6& 09:11:57.38 & $-$51:14:25.2 & 6.7 & 17.8 & 2.9 & 2.2 \\ 
 &&  & & 6.0& 09:11:57.39 & $-$51:14:25.0 & 18.2 & 27.1 & 4.2 & 2.8 \\ 
G274.2126$-$01.1143 && 0.9 & I/MP& 6.0& 09:25:32.50 & $-$52:06:47.0 & 1.3 & 6.6 & 6.5 & 4.9 \\ 
G284.7306+00.3213 && 8.2 & SH& 6.0& 10:29:28.99 & $-$57:26:47.4 & 21.4 & 1,350 & 20.9 & 11.0 \\ 
G285.2611$-$00.0492 && 11.7 & U& 6.0& 10:31:29.02 & $-$58:02:23.8 & 87.1 & 292.0 & 5.8 & 3.2 \\ 
G289.0823$-$00.4107 &A& 10.7 & U& 3.6& 10:56:26.95 & $-$60:09:22.6 & 6.9 & 14.5 & 2.5 & 1.9 \\ 
 &&  & & 6.0& 10:56:26.97 & $-$60:09:22.8 & 11.5 & 15.3 & 3.6 & 2.3 \\ 
 &B& 11 & U& 3.6& 10:56:26.40 & $-$60:09:01.3 & 16.8 & 33.4 & 2.3 & 1.9 \\
\end{longtable}

%% file: radio_table8_unmatched.tex
\begin{longtable}{lcrccccrrrrr}
\caption{\label{tbl:unmatched_parameters}  Observational parameters for unmatched radio sources.} \\ 
\hline
\hline
 & & & & &\multicolumn{2}{c}{Source position}& Peak&	Integ. &\multicolumn{2}{c}{Source} \\ 
Field name& & Offset &Morph.&$\lambda$&$\alpha$ (J2000)  &$\delta$ (J2000)&S$_\nu$&S$_\nu$ & \multicolumn{2}{c}{Diameter } \\ 
&&(\arcsec)&&(cm)&(h m s)&($\degr$~\arcmin\ \arcsec\ ) & (mJy/b)&(mJy)& Max (\arcsec) &Min (\arcsec) \\ 
\hline
\endfirsthead
\caption{continued.} \\ 
\hline
\hline
 & & & &\multicolumn{2}{c}{Source position}& Peak&	Integ. &\multicolumn{2}{c}{Source} \\ 
Field name& &Offset &Morph.&$\lambda$&$\alpha$ (J2000)  &$\delta$ (J2000)&S$_\nu$&S$_\nu$ & \multicolumn{2}{c}{Diameter } \\ 
&&(\arcsec)&&(cm)&(h m s)&($\degr$~\arcmin\ \arcsec\ ) & (mJy/b)&(mJy)& Max (\arcsec) &Min (\arcsec) \\ 
\hline
\endhead
\hline
\endfoot
G254.0548$-$00.0961 && 204.1 & U& 6.0& 08:18:00.74 & $-$35:55:45.9 & 6.3 & 10.0 & 5.4 & 2.4 \\ 
G259.5625$-$00.9213 && 80.6 & U& 3.6& 08:30:33.61 & $-$40:53:19.5 & 12.0 & 13.2 & 2.3 & 1.3 \\ 
G260.6877$-$01.3930 && 286.3 & U& 6.0& 08:31:48.51 & $-$42:06:08.8 & 4.1 & 7.6 & 4.5 & 2.8 \\ 
G263.7434+00.1161 && 115.6 & U& 3.6& 08:48:59.00 & $-$43:32:59.6 & 6.7 & 7.6 & 2.1 & 1.4 \\ 
 &&  & & 6.0& 08:48:58.98 & $-$43:32:59.4 & 8.4 & 12.7 & 4.6 & 2.4 \\ 
G265.1591$-$02.2284 && 135.8 & U& 6.0& 08:43:34.37 & $-$46:08:39.7 & 43.5 & 55.4 & 5.1 & 1.5 \\ 
G269.2580$-$01.5184 && 110.5 & U& 6.0& 09:02:18.21 & $-$48:47:11.3 & 1.4 & 1.5 & 5.3 & 1.7 \\ 
G269.8539$-$00.0630 && 174.5 & S& 6.0& 09:11:25.55 & $-$48:15:27.0 & 48.7 & 81.6 & 3.7 & 3.3 \\
\end{longtable}

%% file: rms_atca_radio_new_v4.bbl
\begin{thebibliography}{27}
\expandafter\ifx\csname natexlab\endcsname\relax\def\natexlab#1{#1}\fi

\bibitem[{{Benjamin} {et~al.}(2005){Benjamin}, {Churchwell}, {Babler},
  {Indebetouw}, {Meade}, {Whitney}, {Watson}, {Wolfire}, {Wolff}, {Ignace},
  {Bania}, {Bracker}, {Clemens}, {Chomiuk}, {Cohen}, {Dickey}, {Jackson},
  {Kobulnicky}, {Mercer}, {Mathis}, {Stolovy}, \& {Uzpen}}]{benjamin2005}
{Benjamin}, R.~A., {Churchwell}, E., {Babler}, B.~L., {et~al.} 2005, \apjl,
  630, L149

\bibitem[{{Busfield} {et~al.}(2006){Busfield}, {Purcell}, {Hoare}, {Lumsden},
  {Moore}, \& {Oudmaijer}}]{busfield2006}
{Busfield}, A.~L., {Purcell}, C.~R., {Hoare}, M.~G., {et~al.} 2006, \mnras,
  366, 1096

\bibitem[{{Clarke} {et~al.}(2006){Clarke}, {Lumsden}, {Oudmaijer}, {Busfield},
  {Hoare}, {Moore}, {Sheret}, \& {Urquhart}}]{clarke2006}
{Clarke}, A.~J., {Lumsden}, S.~L., {Oudmaijer}, R.~D., {et~al.} 2006, ArXiv
  Astrophysics e-prints

\bibitem[{{Cutri} {et~al.}(2003){Cutri}, {Skrutskie}, {van Dyk}, {Beichman},
  {Carpenter}, {Chester}, {Cambresy}, {Evans}, {Fowler}, {Gizis}, {Howard},
  {Huchra}, {Jarrett}, {Kopan}, {Kirkpatrick}, {Light}, {Marsh}, {McCallon},
  {Schneider}, {Stiening}, {Sykes}, {Weinberg}, {Wheaton}, {Wheelock}, \&
  {Zacarias}}]{cutri2003}
{Cutri}, R.~M., {Skrutskie}, M.~F., {van Dyk}, S., {et~al.} 2003, {2MASS All
  Sky Catalog of point sources.} (The IRSA 2MASS All-Sky Point Source Catalog,
  NASA/IPAC Infrared Science
  Archive.~http://irsa.ipac.caltech.edu/applications/Gator/)

\bibitem[{{De Buizer} {et~al.}(2002){De Buizer}, {Watson}, {Radomski},
  {Pi{\~n}a}, \& {Telesco}}]{de_buizer2002}
{De Buizer}, J.~M., {Watson}, A.~M., {Radomski}, J.~T., {Pi{\~n}a}, R.~K., \&
  {Telesco}, C.~M. 2002, \apjl, 564, L101

\bibitem[{{De Pree} {et~al.}(2005){De Pree}, {Wilner}, {Deblasio}, {Mercer}, \&
  {Davis}}]{de_pree2005}
{De Pree}, C.~G., {Wilner}, D.~J., {Deblasio}, J., {Mercer}, A.~J., \& {Davis},
  L.~E. 2005, \apjl, 624, L101

\bibitem[{{Egan} {et~al.}(2003){Egan}, {Price}, {Kraemer}, {Mizuno}, {Carey},
  {Wright}, {Engelke}, {Cohen}, \& {Gugliotti}}]{egan2003}
{Egan}, M.~P., {Price}, S.~D., {Kraemer}, K.~E., {et~al.} 2003, VizieR Online
  Data Catalog, 5114, 0

\bibitem[{{Egan} {et~al.}(1999){Egan}, {Price}, {Moshir}, {Cohen}, \&
  {Tedesco}}]{egan1999}
{Egan}, M.~P., {Price}, S.~D., {Moshir}, M.~M., {Cohen}, M., \& {Tedesco}, E.
  1999, NASA STI/Recon Technical Report N, 14854

\bibitem[{{Ellingsen} {et~al.}(2005){Ellingsen}, {Shabala}, \&
  {Kurtz}}]{ellingsen2005}
{Ellingsen}, S.~P., {Shabala}, S.~S., \& {Kurtz}, S.~E. 2005, \mnras, 357, 1003

\bibitem[{{Forster} \& {Caswell}(2000)}]{forster2000}
{Forster}, J.~R. \& {Caswell}, J.~L. 2000, \apj, 530, 371

\bibitem[{{Giveon} {et~al.}(2005{\natexlab{a}}){Giveon}, {Becker}, {Helfand},
  \& {White}}]{giveon2005a}
{Giveon}, U., {Becker}, R.~H., {Helfand}, D.~J., \& {White}, R.~L.
  2005{\natexlab{a}}, \aj, 129, 348

\bibitem[{{Giveon} {et~al.}(2005{\natexlab{b}}){Giveon}, {Becker}, {Helfand},
  \& {White}}]{giveon2005b}
{Giveon}, U., {Becker}, R.~H., {Helfand}, D.~J., \& {White}, R.~L.
  2005{\natexlab{b}}, \aj, 130, 156

\bibitem[{{Henning} {et~al.}(1984){Henning}, {Friedemann}, {Guertler}, \&
  {Dorschner}}]{Henning1984}
{Henning}, T., {Friedemann}, C., {Guertler}, J., \& {Dorschner}, J. 1984,
  Astronomische Nachrichten, 305, 67

\bibitem[{{Hoare}(2002)}]{hoare2002}
{Hoare}, M.~G. 2002, in ASP Conf. Ser. 267: Hot Star Workshop III: The Earliest
  Phases of Massive Star Birth, ed. P.~{Crowther}, 137--+

\bibitem[{{Hoare} {et~al.}(2005){Hoare}, {Lumsden}, {Oudmaijer}, {Urquhart},
  {Busfield}, {Sheret}, {Clarke}, {Moore}, {Allsopp}, {Burton}, {Purcell},
  {Jiang}, \& {Wang}}]{hoare2005}
{Hoare}, M.~G., {Lumsden}, S.~L., {Oudmaijer}, R.~D., {et~al.} 2005, in IAU
  Symposium, ed. R.~{Cesaroni}, M.~{Felli}, E.~{Churchwell}, \& M.~{Walmsley},
  370--375

\bibitem[{{Kurtz} {et~al.}(1994){Kurtz}, {Churchwell}, {Wood}, \&
  {Myers}}]{kurtz1994}
{Kurtz}, S., {Churchwell}, E., {Wood}, D.~O.~S., \& {Myers}, P. 1994, Bulletin
  of the American Astronomical Society, 26, 907

\bibitem[{{Lumsden} {et~al.}(2002){Lumsden}, {Hoare}, {Oudmaijer}, \&
  {Richards}}]{lumsden2002}
{Lumsden}, S.~L., {Hoare}, M.~G., {Oudmaijer}, R.~D., \& {Richards}, D. 2002,
  \mnras, 336, 621

\bibitem[{{Molinari} {et~al.}(1996){Molinari}, {Brand}, {Cesaroni}, \&
  {Palla}}]{molinari1996}
{Molinari}, S., {Brand}, J., {Cesaroni}, R., \& {Palla}, F. 1996, \aap, 308,
  573

\bibitem[{{Sault} {et~al.}(1995){Sault}, {Teuben}, \& {Wright}}]{sault1995}
{Sault}, R.~J., {Teuben}, P.~J., \& {Wright}, M.~C.~H. 1995, in ASP Conf. Ser.
  77: Astronomical Data Analysis Software and Systems IV, ed. R.~A. {Shaw},
  H.~E. {Payne}, \& J.~J.~E. {Hayes}, 433--+

\bibitem[{{Sridharan} {et~al.}(2002){Sridharan}, {Beuther}, {Schilke},
  {Menten}, \& {Wyrowski}}]{sridharan2002}
{Sridharan}, T.~K., {Beuther}, H., {Schilke}, P., {Menten}, K.~M., \&
  {Wyrowski}, F. 2002, \apj, 566, 931

\bibitem[{{Taylor} {et~al.}(1999){Taylor}, {Carilli}, \& {Perley}}]{vla1999}
{Taylor}, G.~B., {Carilli}, C.~L., \& {Perley}, R.~A., eds. 1999, {Synthesis
  Imaging in Radio Astronomy II}

\bibitem[{{Thompson} {et~al.}(2006){Thompson}, {Hatchell}, {Walsh},
  {MacDonald}, \& {Millar}}]{thompson2006}
{Thompson}, M.~A., {Hatchell}, J., {Walsh}, A.~J., {MacDonald}, G.~H., \&
  {Millar}, T.~J. 2006, \aap, 453, 1003

\bibitem[{{Walsh} {et~al.}(1998){Walsh}, {Burton}, {Hyland}, \&
  {Robinson}}]{walsh1998}
{Walsh}, A.~J., {Burton}, M.~G., {Hyland}, A.~R., \& {Robinson}, G. 1998,
  \mnras, 301, 640

\bibitem[{{Walsh} {et~al.}(1997){Walsh}, {Hyland}, {Robinson}, \&
  {Burton}}]{walsh1997}
{Walsh}, A.~J., {Hyland}, A.~R., {Robinson}, G., \& {Burton}, M.~G. 1997,
  \mnras, 291, 261

\bibitem[{{Wood} \& {Churchwell}(1989)}]{wood1989a}
{Wood}, D.~O.~S. \& {Churchwell}, E. 1989, \apj, 340, 265

\bibitem[{{Wu} {et~al.}(2004){Wu}, {Wei}, {Zhao}, {Shi}, {Yu}, {Qin}, \&
  {Huang}}]{wu2004}
{Wu}, Y., {Wei}, Y., {Zhao}, M., {et~al.} 2004, \aap, 426, 503

\bibitem[{{Wynn-Williams}(1982)}]{wynn-williams1982}
{Wynn-Williams}, C.~G. 1982, \araa, 20, 587

\end{thebibliography}
